\documentclass[11pt]{article}
\usepackage[utf8]{inputenc}

\usepackage{lipsum}
\usepackage{verbatim}
\usepackage{caption}
\usepackage{amssymb}
\usepackage{amsmath}
\usepackage{tabularx}
\usepackage{color}
\usepackage{stmaryrd}
\usepackage{multirow}
\usepackage{mathtools}
\usepackage{epsfig,latexsym}
\usepackage{hyperref}
\usepackage{float}

\let\oldequation\equation
\let\oldendequation\endequation

\usepackage{amsmath,amssymb,amsthm,graphicx,verbatim,bm,fancyhdr,bbm,enumerate,color, multirow, tabularx,mathtools}
\usepackage{hyperref}

\DeclareMathAlphabet{\mathpzc}{OT1}{pzc}{m}{it}

\theoremstyle{definition}
\theoremstyle{plain}

\newtheorem*{proposition*}{Proposition}

\theoremstyle{remark}

\newtheorem*{remark*}{Remark}

\newtheorem*{terminology*}{Terminology}
\newtheorem*{notation*}{Notation}

\newcommand{\E}{\mathbb{E}}

\newcommand{\inquotes}[1]{``#1''}

\usepackage{lineno}

\let\oldalign\align
\let\oldendalign\endalign

\renewenvironment{align}
  {\linenomathNonumbers\oldalign}
  {\oldendalign\endlinenomath}

\let\oldequation\equation
\let\oldendequation\endequation

\renewenvironment{equation}
  {\linenomathNonumbers\oldequation}
  {\oldendequation\endlinenomath}
\title{
Sharp transitions between optimal therapy regimes reveal robust dosing strategies against drug-induced persistence
}

\author{Einar Bjarki Gunnarsson}
\date{%
    \footnotesize Division of Applied Mathematics, Science Institute, University of Iceland \\[0pt]
    }

\renewenvironment{abstract}
 {\small
  \begin{center}
  \bfseries \abstractname\vspace{0pt}\vspace{0pt}
  \end{center}
  \list{}{%
    \setlength{\leftmargin}{21.5mm}% <---------- CHANGE HERE
    \setlength{\rightmargin}{\leftmargin}%
  }%
  \item\relax}
 {\endlist}

\begin{document}

\maketitle

\begin{abstract}
Anti-cancer drug treatment can induce tumor cells to adopt reversible drug-tolerant persister states, facilitating resistance evolution and eventual treatment failure. Although low‑dose and intermittent therapies have been proposed as alternatives to maximum tolerated dose (MTD) therapy in this setting, it remains unclear when and why these strategies are optimal. To address this question, we develop a mathematical model of drug-induced persistence where the drug can promote switching to a drug-tolerant state or inhibit reversions to sensitivity in a dose-dependent manner. We uncover sharp transitions between qualitatively distinct therapy regimes, including intermittent high-dose and continuous low-dose therapy, driven by the relative dose sensitivities of cell death and phenotypic switching. In particular, intermittent treatment is optimal when the drug inhibits reversions at doses too low to effectively kill the sensitive cells. We finally exploit the simple regime structure of optimal therapies to show that robust alternatives to MTD therapy can be identified under incomplete knowledge of the underlying dynamics.
\end{abstract}

\section{Introduction}

Anti-cancer drug treatment commonly fails due to the evolution of resistance \cite{foo2009evolution}. Emerging evidence indicates that drug-tolerant persister cells (DTPs) play an important role in resistance evolution 
in many cancers \cite{sharma2010chromatin,pu2023drug,russo2024cancer,he2024drug,wang2025drug,gunnarsson2020understanding}. DTPs are usually quiescent or slow-cycling, they enable the tumor to survive the initial stages of therapy, and they form a reservoir for the evolution of more permanent resistance mechanisms, both of the genetic and epigenetic kind \cite{sharma2010chromatin,Engelman2016,ramirez2016diverse,shaffer2017rare,russo2022modified}. DTPs have been extensively studied in bacterial populations, 
where reversible phenotypic switching between drug-sensitive and drug-tolerant states can generate drug-tolerant subpopulations even in the absence of antibiotic exposure
\cite{balaban2004bacterial,dhar2007microbial,balaban2013problem,van2017formation}. In this case, the DTPs pre-exist drug treatment as a form of population-level bet-hedging against environmental fluctuations, and they undergo Darwinian selection once the drug is applied \cite{gaal2010exact, muller2013bet, nichol2016stochasticity, jolly2018phenotypic}. In cancer, there is growing evidence that anti-cancer drugs can directly induce tumor cells to adopt reversible drug-tolerant states \cite{pisco2013non,goldman2015temporally,su2017single,russo2022modified},  more consistent with a Lamarckian form of therapy-induced adaptation \cite{huang2012tumor,pisco2013non,alvarez2019interplay,russo2024genetic}. The extent to which DTPs pre-exist treatment or emerge through stress-induced adaptation remains an active area of investigation in both cancer and bacteria \cite{balaban2013problem,cohen2013microbial,van2017formation,russo2024genetic}.

Russo et al.~\cite{russo2022modified} recently combined mathematical modeling with {\em in vitro} experiments to show that targeted drug therapies can induce DTPs in colorectal cancer. They investigated two cell lines, one BRAF V600E-mutated and the other RAS/RAF wild-type, and showed that the dose-dependent dynamics of drug-induced persistence differed substantially between the cell lines. 
In the clinic, genetic and epigenetic differences between patients lead to diverse patient-specific disease dynamics, even for cancers originating from the same organ \cite{felipe2013cancer,vogelstein2013cancer,weinstein2013cancer}. 
This has motivated the development of precision medicine, where the goal is to tailor treatment to an individual patient \cite{garraway2013precision,schork2015personalized,ashley2016towards}.
Mathematical models naturally support precision medicine by providing
a quantitative framework for describing disease dynamics that can be individualized through patient-specific model parameter estimation \cite{jackson2015patient,powathil2015systems,hamis2019blackboard}.
Individualized mathematical models form \inquotes{digital twins} of the patients that can be used to predict the effect of different possible dosing strategies and identify the optimal strategy \cite{wu2022integrating,craig2023practical,lorenzo2024patient,asghar2025application}. However, a key challenge is that estimating patient-specific parameters usually requires longitudinal data that are difficult to obtain in the clinical setting.

Anti-cancer drugs have historically been administered at the 
{\em maximum tolerated dose} (MTD), the highest dose that avoids serious or life-threatening toxicities \cite{kerbel2004anti,norton1986norton,levit2025totality,murphy2023project}. Mathematical models have shown that when the drug induces tumor cells to adopt drug-tolerant or drug-resistant states, low-dose or intermittent therapies can outperform MTD therapy, but the conditions under which such strategies are optimal remain unclear 
\cite{greene2019mathematical,akhmetzhanov2019modelling,greene2020mathematical,kuosmanen2021drug,angelini2022model,gunnarsson2025,gunnarsson2025optimal,corigliano2025optimal,park2026continuum,wang2025integrating,liang2026epigenetic}. In this work, we develop a mathematical model of drug-induced persistence to investigate when and why intermittent treatment is preferable to continuous treatment and lower doses are preferable to higher doses. On the one hand, our approach lends itself to individualized treatment by prescribing optimal dose sizes and frequencies depending on the exact model parameters. On the other hand, our investigation reveals a structural simplicity in optimal therapies that can be exploited to identify effective alternatives to MTD therapy when the model parameters are not known. Even though our models and results are framed in the context of cancer, they apply equally to bacterial, fungal or other microbial populations where drug-induced persistence may occur \cite{el2024drug}.

\section{Results}

\subsection{Mathematical model of drug-induced persistence distinguishing drug effects on tolerance adoption and reversion}

\begin{figure}
    \centering
    \hspace*{-25pt}
    \vspace*{-24pt}
    \includegraphics[scale=1.1]{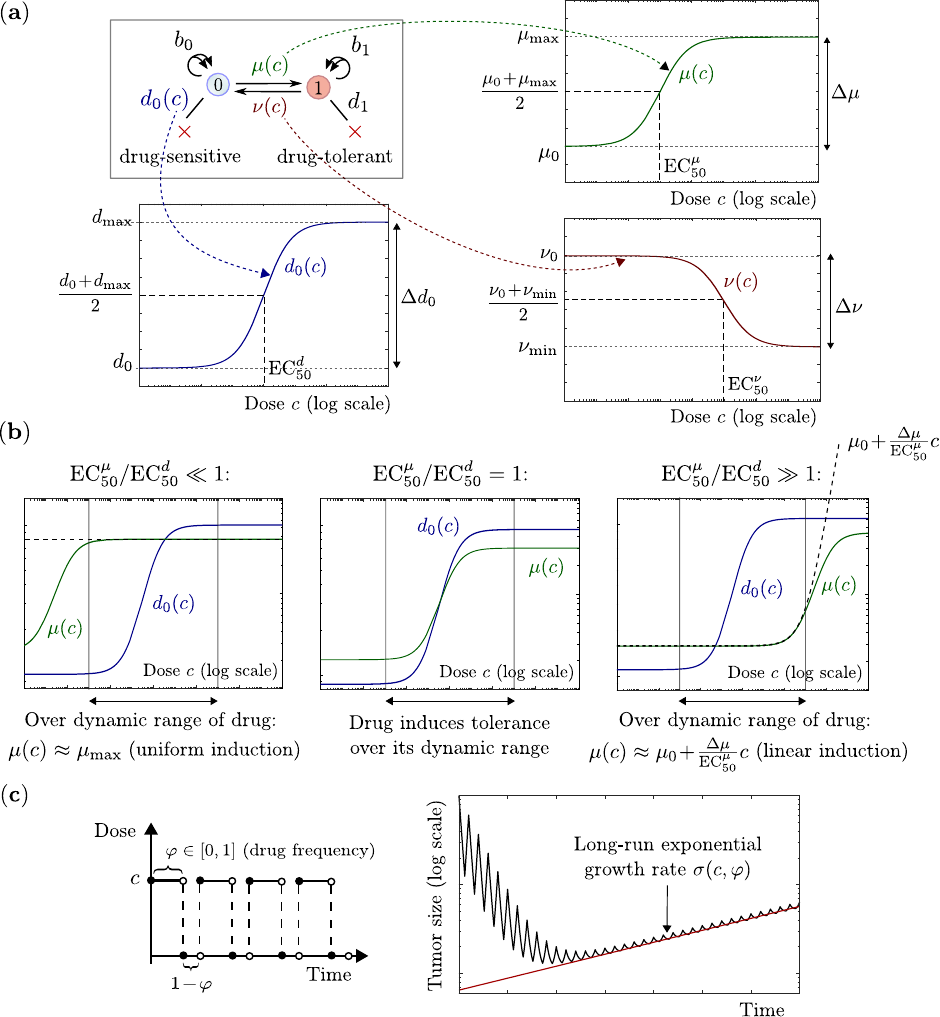}
    \vspace*{6pt}
    \caption{{\bf Mathematical model of drug-induced persistence.} {\bf (a)} Schematic overview of our mathematical model of drug-induced persistence, where cells transition between a drug-sensitive (type-0) and a drug-tolerant (type-1) state. The sensitive cell death rate $d_0(c)$ and the phenotypic transition rates $\mu(c)$ and $\nu(c)$ are assumed to be functions of the drug dose $c$, each following a Michaelis-Menten function. {\bf (b)} The half maximal effective dose ${\rm EC}_{50}^\mu$ for the rate $\mu$ of tolerance adoption is assumed to be distinct from the half maximal effective dose ${\rm EC}_{50}^d$ for the sensitive cell death rate $d_0$. When ${\rm EC}_{50}^\mu/{\rm EC}_{50}^d \ll 1$, the transition rate $\mu(c)$ can be treated as constant over the dynamic range of the drug, and when ${\rm EC}_{50}^\mu/{\rm EC}_{50}^d \gg 1$, $\mu(c)$ can be treated as linear in $c$ over the dynamic range of the drug. {\bf (c)} Under a cyclic dosing strategy $(c,\varphi)$ applying the dose $c$ a proportion $\varphi$ of the time, the evolution of the tumor size is biphasic, where first the drug-sensitive cells are killed and then the tumor grows slowly driven by the drug-tolerant cells. Expression \eqref{eq:sigmapulsed} gives the long-run exponential growth rate of the tumor, indicated by the red line.
    }
    \label{fig1}
\end{figure}

We propose a mathematical model where each tumor cell is able to transition between two states, a drug-sensitive and a drug-tolerant state (Figure \ref{fig1}a). A drug-sensitive cell divides into two cells at rate $b_0$, dies at rate $d_0$, and transitions to the tolerant state at rate $\mu$. A drug-tolerant cell divides at rate $b_1$, dies at rate $d_1$, and reverts to the sensitive state at rate $\nu$. The net division rate is $\lambda_0 = b_0-d_0$ for sensitive cells and $\lambda_1 = b_1-d_1$ for tolerant cells. We assume that the tumor is being treated by a cytotoxic drug, meaning that the drug increases the death rate $d_0$ of drug-sensitive cells. However, our results also apply to cytostatic drugs that reduce the division rate $b_0$ of sensitive cells (Supplementary Note 1). More specifically, we assume that $d_0$ follows a Michaelis-Menten function of the drug dose $c$:
\begin{align} \label{eq:sensitivecelldeathrate}
    d_0(c) = d_0 + (d_{\rm max}-d_0) \cdot \frac{c}{c+{\rm EC}_{50}^d} = d_0 + \Delta d_0 \cdot \frac{c/{\rm EC}_{50}^d}{c/{\rm EC}_{50}^d+1}.
\end{align}
Here, $d_0$ is the sensitive cell death rate in the absence of drug, $d_{\rm max}$ is the death rate under an arbitrarily high drug dose, $\Delta d_0 = d_{\rm max}-d_0$ is the maximal drug effect, and ${\rm EC}_{50}^d$ is the dose at which the drug has half its maximal effect (Figure \ref{fig1}a). We assume an allowable range $[c_{\rm min},c_{\rm max}] = [10^{-3},10^3] \times {\rm EC}_{50}^d$ for the drug dose, below which the drug is ineffective at killing the sensitive cells, and above which there is negligible benefit to increasing the dose. We will refer to this range as the {\em dynamic range} of the drug (Figure \ref{fig1}b). 

Since we are interested in modeling drug-induced persistence, we next assume that the drug can increase the rate $\mu$ at which sensitive cells adopt tolerance, decrease the rate $\nu$ at which tolerant cells revert to sensitivity, or do both simultaneously. In each case, the drug induces the tumor cells to spend a greater proportion of time in the drug-tolerant state than in the drug-sensitive state. The transition rates $\mu(c)$ and $\nu(c)$ are assumed to follow their own Michaelis-Menten functions of the drug dose $c$, where $\mu_0 > 0$ and $\nu_0 > 0$ (Figure \ref{fig1}a):
\begin{align*}
    &\mu(c) = \mu_0 + (\mu_{\rm max}-\mu_0) \cdot \frac{c}{c+{\rm EC}_{50}^\mu} = \mu_0 + \Delta \mu \cdot \frac{c/{\rm EC}_{50}^d}{c/{\rm EC}_{50}^d+{\rm EC}_{50}^\mu/{\rm EC}_{50}^d}, \\
    &\nu(c) = \nu_0 - (\nu_0-\nu_{\rm min}) \cdot \frac{c}{c+{\rm EC}_{50}^\nu} 
= \nu_0 - \Delta \nu \cdot \frac{c/{\rm EC}_{50}^d}{c/{\rm EC}_{50}^d+{\rm EC}_{50}^\nu/{\rm EC}_{50}^d}. 
\end{align*}
A key assumption that distinguishes our model from existing models is that the half maximal effective doses ${\rm EC}_{50}^\mu$ and ${\rm EC}_{50}^\nu$ for $\mu$ and $\nu$ can be distinct from one another, and each can be distinct from the half maximal effective dose ${\rm EC}_{50}^d$ for the sensitive cell death rate $d_0$ (Figure \ref{fig1}b). The final expressions for $\mu(c)$ and $\nu(c)$ show that when ${\rm EC}_{50}^d$ is treated as the unit dose, the ratios ${\rm EC}_{50}^\mu/{\rm EC}_{50}^d$ and ${\rm EC}_{50}^\nu/{\rm EC}_{50}^d$ become the half maximal effective doses. These dimensionless ratios describe the relative dose sensitivities of cell death and phenotypic switching, and we will show that they largely determine whether intermittent treatment outperforms continuous treatment and lower doses outperform higher doses.

Many existing mathematical models of drug-induced genetic or non-genetic resistance assume a linear rate of transition from sensitivity to resistance, $\mu(c) = \mu_0 + kc$ \cite{greene2019mathematical,greene2020mathematical,kuosmanen2021drug,corigliano2025optimal,gunnarsson2025optimal}. Another common assumption is that the transition rate increases uniformly in the presence of drug, meaning that $\mu(0) = \mu_0$ and $\mu(c) = \mu_{\rm max}$ whenever $c>0$ \cite{mathur2022optimal,russo2022modified,corigliano2025optimal, gunnarsson2025optimal}. These two forms of drug-induced adaptation were recently observed experimentally in colorectal cancer \cite{russo2022modified}. Our model allows for more general dynamics that incorporate linear and uniform induction of drug tolerance as extreme cases. More specifically, when ${\rm EC}_{50}^\mu/{\rm EC}_{50}^d \gg 1$, meaning that the tolerance-inducing effects of the drug occur at higher doses than are required for cell kill, the transition rate $\mu(c)$ can be approximated as linear over the dynamic range of the drug (Figure \ref{fig1}b, Methods). Conversely, when ${\rm EC}_{50}^\mu/{\rm EC}_{50}^d \ll 1$, $\mu(c)$ can be approximated as constant over the dynamic range of the drug (Figure \ref{fig1}b, Methods). We will pay special attention to these two extreme cases, as well as the case ${\rm EC}_{50}^\mu/{\rm EC}_{50}^d=1$, where the tolerance-inducing effects of the drug coincide with its cell-killing effects. 

In our investigation of optimal dosing strategies, we will focus on cyclic strategies where a drug dose $c \in [c_{\rm min},c_{\rm max}]$ is applied $\varphi \in [0,1]$ proportion of the time (Figure \ref{fig1}c). 
The evolution of the tumor under a cyclic strategy is biphasic, where first the tumor decays rapidly as the drug-sensitive cells are killed, and then it grows slowly in the long run driven by the slow-cycling drug-tolerant cells.
Under the assumption that the length of each treatment cycle is arbitrarily short, the long-run tumor growth rate is given by
\begin{align} \label{eq:sigmapulsed}
    \sigma(c,\varphi) :=  \frac12 \Big(&(\bar\lambda_0(c,\varphi)-\bar\mu(c,\varphi)) + (\lambda_1-\bar\nu(c,\varphi)) \nonumber \\
    &+ \sqrt{((\bar\lambda_0(c,\varphi)-\bar\mu(c,\varphi))-(\lambda_1-\bar\nu(c,\varphi)))^2+4\bar\mu(c,\varphi)\bar\nu(c,\varphi)}\Big),
\end{align}
where $\bar\lambda_0(c,\varphi)$, $\bar\mu(c,\varphi)$ and $\bar\nu(c,\varphi)$ are weighted averages of the corresponding rates over a treatment cycle (Methods). 
This simplifying assumption is motivated by our previous work, where we showed that  for a cyclic strategy with a given cycle length, there exists a strategy with half the cycle length that leads to a smaller or equal long-run growth rate \cite{gunnarsson2025optimal}.
We are interested in finding the cyclic strategy that leads to the lowest tumor size in the long term, which corresponds to finding the strategy $(c^\ast,\varphi^\ast)$ that minimizes the long-run growth rate in \eqref{eq:sigmapulsed}. 
We are also interested in comparing the optimal therapy with continuous maximum dosing, which we refer to as {\em MTD therapy} for simplicity. We acknowledge that the maximum allowable dose $c_{\rm max}$ in our mathematical model may differ from the MTD in a clinical context, and that the MTD should ideally be defined on a patient-specific basis.

\begin{figure}
\vspace*{-40pt}
    \centering
    \includegraphics[scale=1.1]{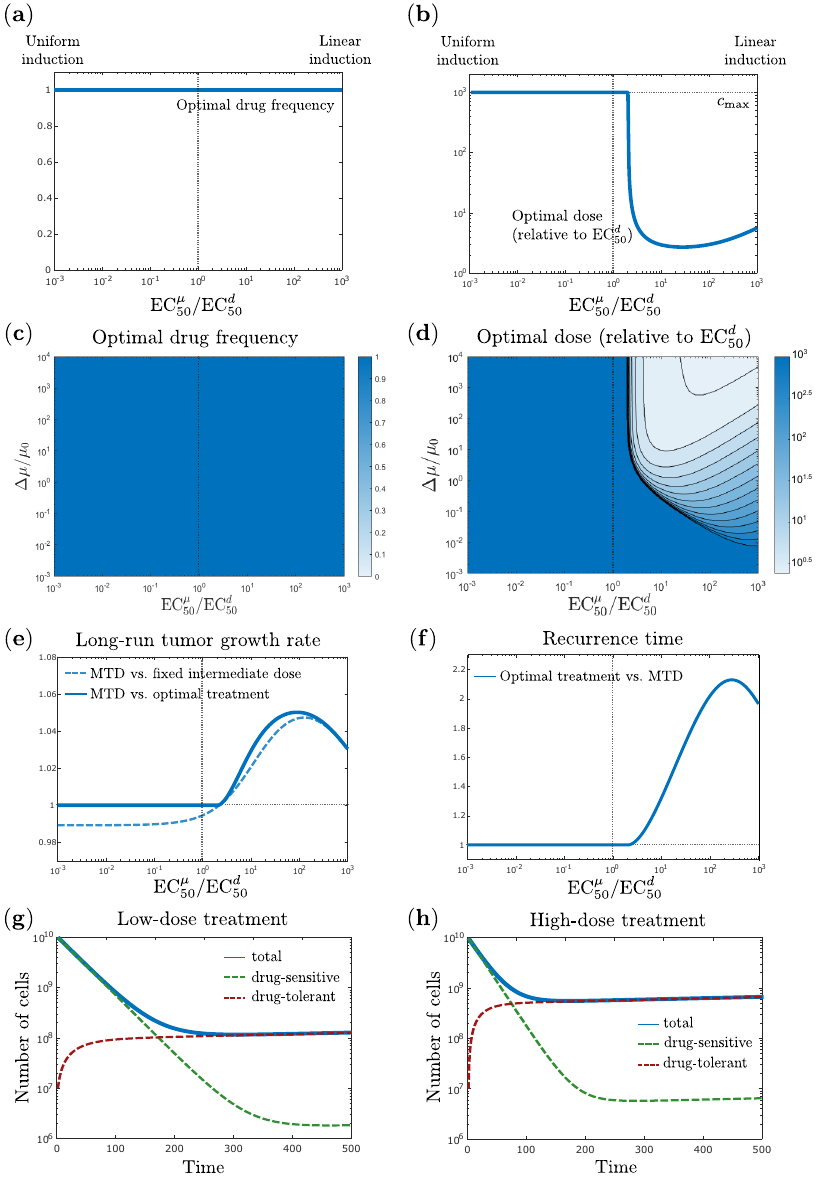}
    \vspace*{-6pt}
    \caption{
    {\bf Optimal therapy when the drug only affects forward transition rate from sensitivity to tolerance.}
    {\bf (a)} Optimal drug frequency and {\bf (b)} optimal dose size as a function of ${\rm EC}_{50}^\mu/{\rm EC}_{50}^d$ for maximal drug effect $\Delta\mu = 10^2 \times \mu_0$. {\bf (c)} Optimal drug frequency and {\bf (d)} optimal dose size as a function of ${\rm EC}_{50}^\mu/{\rm EC}_{50}^d$ and $\Delta \mu/\mu_0$.
    {\bf (e)} Ratio between the long-run tumor growth rate under MTD therapy vs.~the optimal treatment (solid curve) and the constant-dose treatment with dose size $c = 10 \times {\rm EC}_{50}^d$ (dashed curve).
    {\bf (f)} Ratio between the recurrence time $\tau$ under the optimal treatment vs.~MTD therapy.
    {\bf (g)} Sizes of drug-sensitive and drug-tolerant subpopulations (dashed curves) and overall tumor size (solid curve) under constant low-dose treatment ($c = 5 \times {\rm EC}_{50}^d$) when ${\rm EC}_{50}^\mu / {\rm EC}_{50}^d = 100$.
    {\bf (h)} Same as {\bf (g)} but now under constant high-dose treatment ($c = 100 \times {\rm EC}_{50}^d$).
    }
    \label{fig2}
\end{figure}

\subsection{Continuous low-dose or high-dose treatment is optimal when the drug only increases the rate of tolerance adoption}

\noindent
We first consider the case where the drug only increases the transition rate $\mu$ from sensitivity to tolerance, without inhibiting reversion to sensitivity ($\Delta \nu = 0$). All results are shown in a parameter regime inspired by {\em in vitro} non-small cell lung cancer experiments, where the sensitive cells proliferate rapidly in the absence of drug and die rapidly in its presence, and the tolerant cells proliferate slowly both on and off drug (Methods). 

We find that it is optimal to apply the drug continuously for all values of ${\rm EC}_{50}^\mu/{\rm EC}_{50}^d$ (Figure \ref{fig2}a). Furthermore, it is optimal to give the maximum dose $c_{\rm max}$ when ${\rm EC}_{50}^\mu / {\rm EC}_{50}^d \ll 1$ and a low dose when ${\rm EC}_{50}^\mu / {\rm EC}_{50}^d \gg 1$, with a relatively sharp transition between the two regimes (Figure \ref{fig2}b). To understand why, note that when ${\rm EC}_{50}^\mu / {\rm EC}_{50}^d \ll 1$, there is no downside in giving the maximum dose, since the transition rate $\mu(c)$ can be treated as constant over the dynamic range of the drug. When ${\rm EC}_{50}^\mu / {\rm EC}_{50}^d \gg 1$, $\mu(c)$ increases linearly over the dynamic range of the drug. In this case, a lower dose is selected that simultaneously kills sensitive cells and moderates tolerance induction. These insights hold as long as the maximal drug effect $\Delta \mu = \mu_{\rm max}-\mu_0$ is sufficiently large compared to the baseline rate $\mu_0$. As $\Delta \mu$ decreases, the transition between the high-dose and low-dose regimes becomes less sharp, and when $\Delta \mu$ is small, MTD therapy is optimal for all values of ${\rm EC}_{50}^\mu/{\rm EC}_{50}^d$ (Figure \ref{fig2}c,d).

Despite the relatively sharp transition between the high-dose and low-dose regimes, the optimal low dose for ${\rm EC}_{50}^\mu / {\rm EC}_{50}^d \gg 1$
produces a similar long-run growth rate to the maximum dose (Figure \ref{fig2}e, solid curve). In fact, we can select a single intermediate dose ($10 \times {\rm EC}_{50}^d$) that performs similarly both to the maximum dose when ${\rm EC}_{50}^\mu / {\rm EC}_{50}^d \ll 1$ and to the optimal low dose when ${\rm EC}_{50}^\mu / {\rm EC}_{50}^d \gg 1$ (Figure \ref{fig2}e, dashed curve). This indicates that the long-run growth rate is relatively insensitive to the dose size, although it becomes increasingly sensitive to the dose size as the maximal drug effect $\Delta \mu$ increases (Figure \ref{fig:s1}). The benefit of applying a low dose becomes more apparent when the recurrence time $\tau$, the time it takes the tumor to return to its original size, is considered (Figure \ref{fig2}f, Methods). This is due to the biphasic nature of tumor evolution under treatment. Under a low dose, the drug kills the sensitive cells gradually, while simultaneously inducing a small proportion of them to adopt tolerance, after which the tolerant subpopulation grows slowly (Figure \ref{fig2}g). Under a high dose, the sensitive cells are killed more rapidly, but the drug induces a larger number of them to adopt tolerance at the beginning. This leads to a larger tumor size and a shorter recurrence time, even though the long-run growth rate is similar to the low-dose case (Figure \ref{fig2}h).

These results generalize the findings of our previous work \cite{gunnarsson2025optimal}, where we showed that low continuous dosing is 
optimal under linear induction of tolerance (${\rm EC}_{50}^\mu / {\rm EC}_{50}^d \gg 1$) and continuous maximum dosing is optimal under uniform induction (${\rm EC}_{50}^\mu / {\rm EC}_{50}^d \ll 1$). The present work reveals a bifurcation between the low-dose and high-dose regimes across the entire range of possible values of ${\rm EC}_{50}^\mu/{\rm EC}_{50}^d$. Interestingly, our results also show that when ${\rm EC}_{50}^\mu / {\rm EC}_{50}^d$ = 1, meaning that the tolerance-inducing effects of the drug coincide with its cell-killing effects, it is optimal to apply MTD therapy. In other words, for low-dose therapy to be optimal, the drug must induce tolerance at higher doses than are required to kill the sensitive cells. At the same time, our results show that irrespective of the value of ${\rm EC}_{50}^\mu / {\rm EC}_{50}^d$, it is possible to achieve at least comparable results to
MTD therapy using a much lower dose.

\subsection{Continuous low-dose or intermittent high-dose treatment is optimal when the drug only inhibits the rate of reversion}

\begin{figure*}
\vspace*{-18pt}
    \centering
    \includegraphics[scale=1.1]{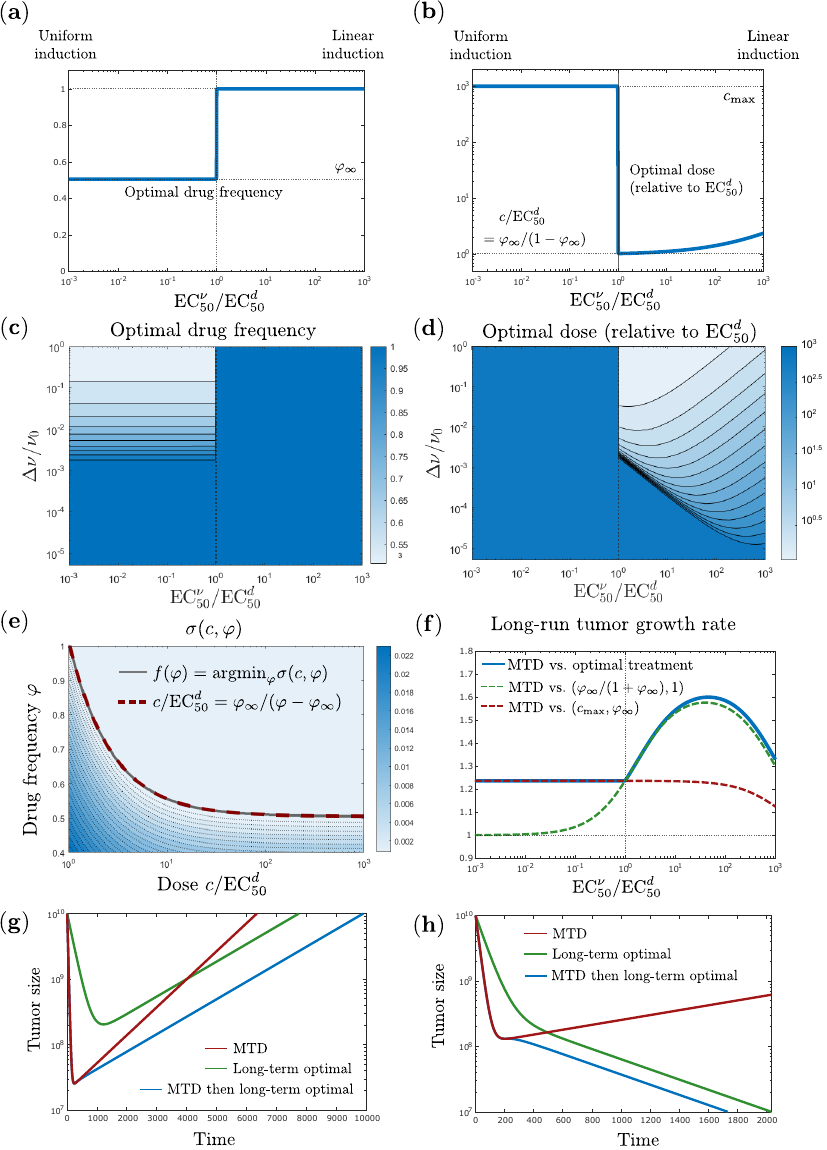}
    \vspace*{-12pt}
\end{figure*}

\begin{figure}
    \caption{
    {\bf Optimal therapy when the drug only affects reversion rate from tolerance to sensitivity.}
    {\bf (a)} Optimal drug frequency and {\bf (b)} optimal dose size as a function of ${\rm EC}_{50}^\nu/{\rm EC}_{50}^d$ for maximal drug effect $\Delta\nu = \nu_0$. {\bf (c)} Optimal drug frequency and {\bf (d)} optimal dose size as a function of ${\rm EC}_{50}^\nu/{\rm EC}_{50}^d$ and $\Delta \nu$.
    {\bf (e)} Contour map showing that the curve $c/{\rm EC}_{50}^d = {\varphi_\infty}/({\varphi-\varphi_\infty})$ (dashed red curve)
    aligns with the level curves of $\sigma(c,\varphi)$,
    indicating that the long-run growth rate is fixed on this curve.
Furthermore, this curve coincides with the optimal drug frequency as a function of the dose (grey curve), indicating that all strategies along this curve minimize the long-run growth rate.
    {\bf (f)} Ratio between the long-run tumor growth rate under MTD therapy vs.~the optimal treatment (solid curve), the constant-dose treatment with dose size $c/{\rm EC}_{50}^d = \varphi_\infty/(1-\varphi_\infty)$ (dashed green curve) and the intermittent maximum-dose treatment with drug frequency $\varphi = \varphi_\infty$ (dashed red curve).
    {\bf (g)} Tumor size evolution under MTD therapy (red curve), the long-term optimal therapy (green curve), and a therapy giving the maximum dose until the tumor starts rebounding before switching to the optimal long-term therapy (blue curve), when ${\rm EC}_{50}^\nu / {\rm EC}_{50}^d = 50$.
    {\bf (h)} Same as {\bf (g)} where the transition rates have been changed from the baseline values to $\mu_0 = 0.0004$, $\nu_0 = 0.003$ and $\Delta \nu =0.003$. In this parameter regime, it is possible to eradicate the tumor under the optimal therapy, whereas the tumor eventually grows to its initial size under MTD therapy.
    }
    \label{fig3}
\end{figure}

\noindent We next consider the case where the drug only reduces the reversion rate $\nu$ from tolerance to sensitivity ($\Delta \mu = 0$). In this case, we observe a phase transition (abrupt change) in the optimal drug frequency and dose size when ${\rm EC}_{50}^\nu / {\rm EC}_{50}^d = 1$ (Figure \ref{fig3}a,b). For ${\rm EC}_{50}^\nu  / {\rm EC}_{50}^d < 1$, it is optimal to apply the maximum dose intermittently, and the optimal drug frequency is insensitive to the value of ${\rm EC}_{50}^\nu  / {\rm EC}_{50}^d$. For ${\rm EC}_{50}^\nu  / {\rm EC}_{50}^d > 1$, on the other hand, it is optimal to apply a low dose continuously. These insights hold as long as the maximal drug effect $\Delta \nu = \nu_0-\nu_{\rm min}$ is sufficiently large compared to the baseline rate $\nu_0$. As $\Delta \nu$ decreases, the optimal drug frequency for ${\rm EC}_{50}^\nu  / {\rm EC}_{50}^d < 1$ and the optimal dose size for ${\rm EC}_{50}^\nu  / {\rm EC}_{50}^d > 1$ increase, and for small $\Delta \nu$, MTD therapy becomes optimal (Figure \ref{fig3}c,d).

To understand why these optimal therapy regimes emerge, it is useful to consider the extreme cases of uniform (${\rm EC}_{50}^\nu / {\rm EC}_{50}^d \ll 1$) and linear (${\rm EC}_{50}^\nu / {\rm EC}_{50}^d \gg 1$) induction of tolerance. When ${\rm EC}_{50}^\nu / {\rm EC}_{50}^d \ll 1$, the reversion rate $\nu(c)$ can be treated as constant over the dynamic range of the drug. In this case, there is no downside in giving the maximum dose, and it is beneficial to periodically remove the drug, since the tolerant cells revert to sensitivity faster off drug than on drug in this case. When ${\rm EC}_{50}^\nu / {\rm EC}_{50}^d \gg 1$, on the other hand, the reversion rate $\nu(c)$ gradually reduces as the dose $c$ increases. Under a high dose, a tolerant cell reverting back to sensitivity will most likely die soon thereafter, meaning that the reversion rate acts like a death rate. Therefore, as a higher dose is applied, the death rate of tolerant cells decreases and their fitness increases, which is why a lower dose is preferred. The fact that intermittent maximum dosing and continuous low dosing are optimal for uniform and linear induction of tolerance is consistent with our previous work \cite{gunnarsson2025optimal}, but what is more surprising is that these two regimes apply across the entire range of possible values of ${\rm EC}_{50}^\nu / {\rm EC}_{50}^d$.

To further support the numerical evidence for a phase transition at ${\rm EC}_{50}^\nu/{\rm EC}_{50}^d=1$, we conducted a mathematical analysis (Supplementary Note 1). First, we showed that in the limit of an arbitrarily high dose ($c \to \infty$), the optimal drug frequency $\varphi_\infty$ is independent of the value of ${\rm EC}_{50}^\nu / {\rm EC}_{50}^d$, as is further supported by the dotted line in Figure \ref{fig3}a. Second, we showed that when ${\rm EC}_{50}^\nu/{\rm EC}_{50}^d = 1$, the long-run growth rate depends on the dose size and drug frequency only through the combined parameter $x = \varphi c/(c+{\rm EC}_{50}^d)$. Consequently, the optimal strategy is not unique. Instead, all 
cyclic strategies 
$(c,\varphi)$ satisfying
\[
c/{\rm EC}_{50}^d = {\varphi_\infty}/({\varphi-\varphi_\infty})
\]
with $\varphi_\infty < \varphi \leq1$ are optimal, as is further supported by the contour map for the long-run growth rate function $\sigma(c,\varphi)$ in Figure \ref{fig3}e. One of these strategies is the continuous strategy $(c_1,1)$ with dose size $c_1/{\rm EC}_{50}^d = {\varphi_\infty}/(1-\varphi_\infty)$, shown using the dotted line in Figure \ref{fig3}b. 
Finally, we showed that when ${\rm EC}_{50}^\nu/{\rm EC}_{50}^d = 1 + \delta$ and $|\delta|$ is small, then on the curve $c/{\rm EC}_{50}^d = {\varphi_\infty}/({\varphi-\varphi_\infty})$, the long-run growth rate is of the form $\sigma(c,\varphi) = a+b(1-\varphi_\infty/\varphi)\delta$ up to first order in $\delta$, where $a$ and $b<0$ are constants.
This function is strictly increasing in $\varphi$ when $\delta<0$ and strictly decreasing when $\delta>0$, so $\varphi = \varphi_\infty$ or $\varphi=1$ minimizes the long-run growth rate
depending on 
whether
${\rm EC}_{50}^\nu/{\rm EC}_{50}^d$ is slightly smaller or larger than 1.

Together, our computational and mathematical results indicate that the optimal treatment transitions abruptly from an intermittent maximum-dose strategy to a continuous low-dose strategy as ${\rm EC}_{50}^\nu/{\rm EC}_{50}^d$ crosses the value 1, and that these strategies belong to a continuum of optimal treatments when ${\rm EC}_{50}^\nu/{\rm EC}_{50}^d=1$. 
The mathematical analysis extends these insights beyond the parameter regime used in Figure \ref{fig3} to any parameter regime where
$\lambda_0 -  (d_{\rm max}-d_0) \varphi_\infty <\lambda_1$ and $0 < \varphi_\infty < 1$.
The former condition simply says that when an arbitrarily high dose is applied with frequency $\varphi_\infty$, the treatment reduces the time-average net growth rate of sensitive cells below that of tolerant cells.

For all values of ${\rm EC}_{50}^\nu/{\rm EC}_{50}^d$, the optimal strategy leads to a substantially lower long-run growth rate than MTD therapy, especially when ${\rm EC}_{50}^\nu/{\rm EC}_{50}^d>1$ (Figure \ref{fig3}f, blue curve). The continuous strategy $(c_1,1)$ with dose size $c_1/{\rm EC}_{50}^d = \varphi_\infty/(1-\varphi_\infty)$ performs at least as well as MTD therapy for ${\rm EC}_{50}^\nu/{\rm EC}_{50}^d \ll 1$ and almost as well as the optimal strategy for ${\rm EC}_{50}^\nu/{\rm EC}_{50}^d \gg 1$ (Figure \ref{fig3}f, green dashed curve). The intermittent maximum-dose strategy $(\varphi_\infty,c_{\rm max})$ is optimal for ${\rm EC}_{50}^\nu/{\rm EC}_{50}^d < 1$ and performs better than MTD therapy for ${\rm EC}_{50}^\nu/{\rm EC}_{50}^d > 1$ (Figure \ref{fig3}f, red dashed curve). This suggests that when the value of ${\rm EC}_{50}^\nu/{\rm EC}_{50}^d$ is not known, and the optimal strategy cannot be identified, it may still be possible to select a specific dose-frequency combination that outperforms MTD therapy across all possible values of ${\rm EC}_{50}^\nu/{\rm EC}_{50}^d$. We explore this further in the final results section.

In the previous section, the long-run tumor growth rate was relatively insensitive to the dose size, and the benefit of applying a low dose was most apparent during the initial stages of treatment. Here, the opposite is true. First, there is no downside in applying a high dose at the beginning, since the drug does not induce the sensitive cells to adopt tolerance. On the other hand, a high dose increases the fitness of tolerant cells, which increases the long-run tumor growth rate. Thus, by starting with the maximum dose and then switching to the optimal long-term therapy once the tumor starts rebounding, it is possible to significantly extend the tumor recurrence time (Figure \ref{fig3}g). More generally, while our focus in this work is on finding the cyclic strategy that leads to the lowest long-run growth rate, it may be possible to improve upon this strategy by adopting a different strategy at the beginning. We refer to our previous work \cite{gunnarsson2025optimal} for a further discussion. We also note that while we generally assume in this work that $\lambda_1 - \nu_0 > 0$, so the tolerant cell population grows slowly over time under treatment, the benefits of applying the optimal strategy can become even more pronounced in parameter regimes where $\lambda_1 - \nu_0 < 0$.
In that case, applying the optimal strategy instead of MTD therapy can turn tumor recurrence into tumor eradication (Figure \ref{fig3}h).

\subsection{Inhibition of reversion controls optimal drug frequency and induction of tolerance controls optimal dose}

\begin{figure}
    \centering
    \hspace*{-20pt}
    \includegraphics[scale=0.4]{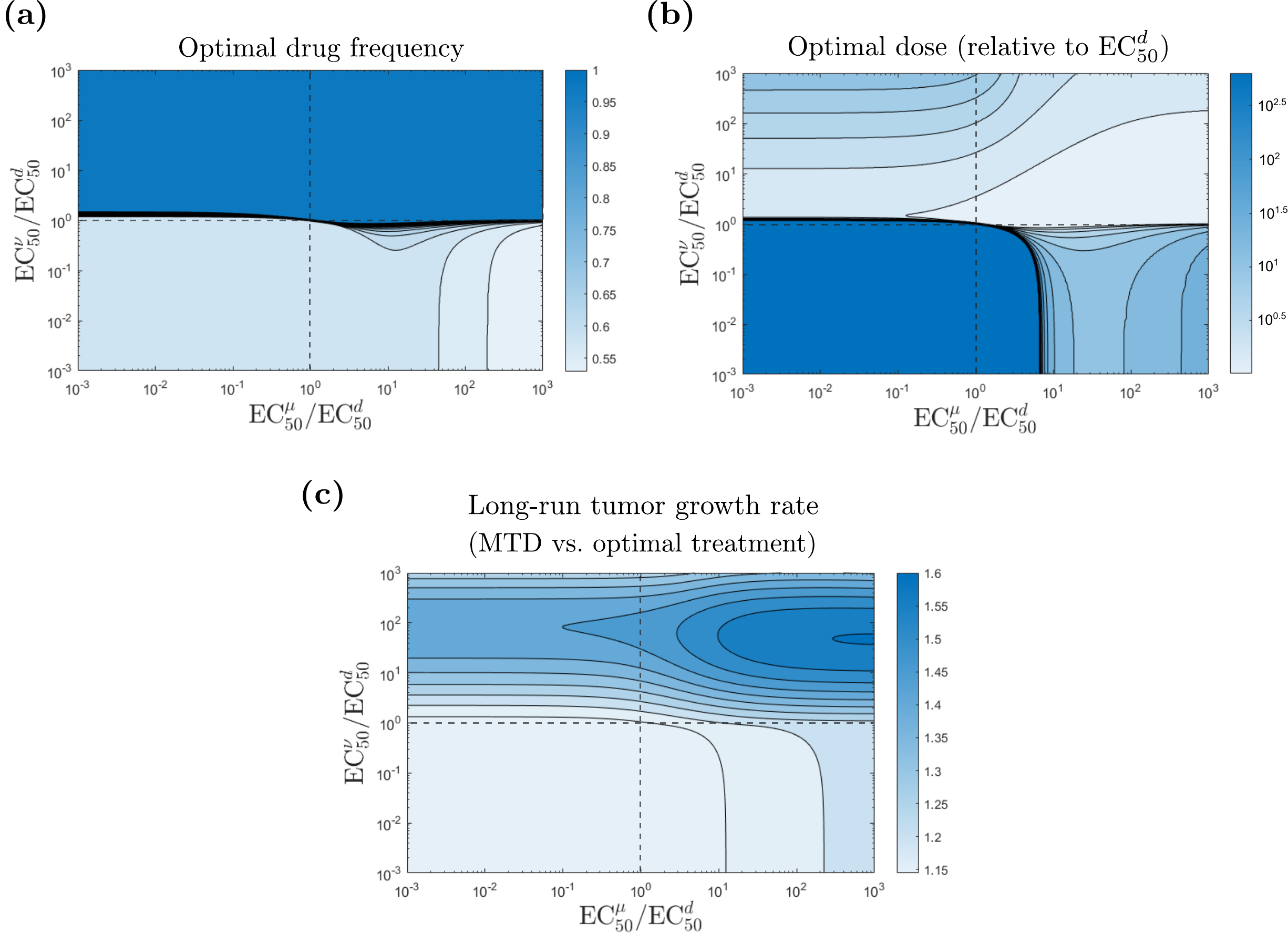}
    \caption{
    {\bf Optimal therapy when the drug affects both transition rates simultaneously.}
    {\bf (a)} Optimal drug frequency and {\bf (b)} optimal dose size as a function of ${\rm EC}_{50}^\mu/{\rm EC}_{50}^d$ and ${\rm EC}_{50}^\nu/{\rm EC}_{50}^d$ when $\Delta \mu = 10^2 \mu_0$ and $\Delta \nu = \nu_0$. {\bf (c)} Ratio between the long-run tumor growth rate under MTD therapy vs.~the optimal treatment as a function of ${\rm EC}_{50}^\mu/{\rm EC}_{50}^d$ and ${\rm EC}_{50}^\nu/{\rm EC}_{50}^d$.
    }
    \label{fig4}
\end{figure}

We next examine to what extent the insights of the previous sections generalize when the drug affects both transition rates simultaneously. We first note that continuous low dosing is optimal when ${\rm EC}_{50}^\mu/{\rm EC}_{50}^d, {\rm EC}_{50}^\nu/{\rm EC}_{50}^d > 1$ (NE quadrants in Figures \ref{fig4}a,b) and intermittent maximum dosing is optimal when ${\rm EC}_{50}^\mu/{\rm EC}_{50}^d, {\rm EC}_{50}^\nu/{\rm EC}_{50}^d < 1$  (SW quadrants in Figures \ref{fig4}a,b), the same two regimes that applied when the drug only affected $\nu$. ${\rm EC}_{50}^\nu/{\rm EC}_{50}^d$ primarily controls whether intermittent treatment is preferable to continuous treatment, and there continues to be a sharp transition between these two regimes, although there is no longer a phase transition precisely at ${\rm EC}_{50}^\nu/{\rm EC}_{50}^d=1$ (Figure \ref{fig4}a). On the other hand, ${\rm EC}_{50}^\mu/{\rm EC}_{50}^d$ primarily controls the size of the optimal dose, with higher values of ${\rm EC}_{50}^\mu/{\rm EC}_{50}^d$ favoring lower doses, consistent with our results for $\mu$ individually (Figure \ref{fig4}b). The optimal strategy leads to a substantially lower long-run growth rate than MTD therapy for all values of ${\rm EC}_{50}^\mu/{\rm EC}_{50}^d$ and  ${\rm EC}_{50}^\nu/{\rm EC}_{50}^d$, especially when ${\rm EC}_{50}^\mu/{\rm EC}_{50}^d,{\rm EC}_{50}^\nu/{\rm EC}_{50}^d > 1$ (Figure \ref{fig4}c). As before, these insights apply whenever the maximal drug effects $\Delta \mu$ and $\Delta \nu$ are sufficiently large compared to the baseline transition rates $\mu_0$ and $\nu_0$. When $\Delta \mu$ or $\Delta \nu$ is sufficiently small, the results mimic what we observed for $\nu$ and $\mu$ individually, respectively (Figure \ref{fig:s2}).

A key takeaway is that in our model, continuous treatment is always optimal when the drug induces sensitive cells to adopt tolerance without inhibiting their reversion to sensitivity.
Intermittent treatment becomes optimal when the drug inhibits reversions to sensitivity at doses that are too low to effectively kill sensitive cells, regardless of whether it also induces tolerance adoption.
We note that several previous studies on intermittent and adaptive therapies have considered models where sensitive cells adopt tolerance (or resistance) only in the presence of drug and tolerant cells revert to sensitivity only in its absence, typically without explicit dose dependence
\cite{yin2022anti,kim2021adaptive,corigliano2025optimal,hamis2026growth}.
In the context of our framework, these dynamics are equivalent to simultaneous induction of tolerance and inhibition of reversion, and our analysis indicates that it is inhibition of reversion, rather than induction of tolerance, that drives the 
preference for intermittent treatment in such models.

Overall, the results of the previous three subsections show that drug-induced persistence is a multifaceted phenomenon that can call for a broad range of dose-frequency combinations depending on the underlying dynamics. At the same time, a few simple therapy regimes have emerged, the transitions between these regimes are surprisingly sharp, and the optimal drug frequencies and dose sizes are relatively stable away from the transition points. 
We explore potential implications of this structural simplicity in the following section.

\subsection{Leveraging the sharp transitions between optimal regimes to derive robust strategies that outperform MTD therapy}

\begin{figure}
\vspace*{-24pt}
    \centering
    \hspace*{-50pt} 
    \includegraphics[scale=0.43]{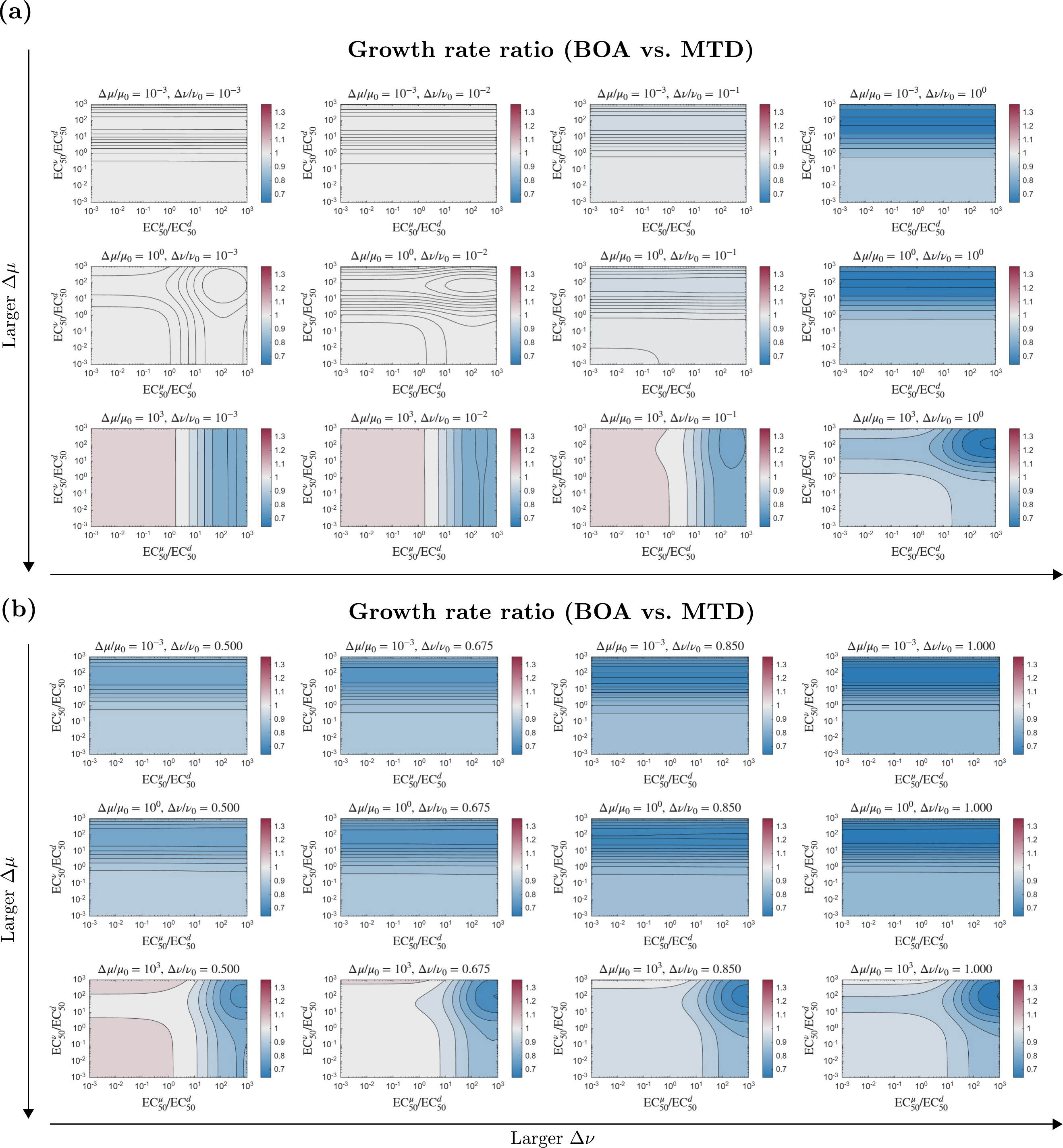}
    \vspace*{-12pt}
    \caption{
    {\bf Benefit of best-on-average (BOA) strategy relative to MTD therapy in a heterogeneous group of virtual patients.}
    {\bf (a)} Each panel shows the ratio between the long-run tumor growth rate under the BOA strategy vs.~MTD therapy as a function of ${\rm EC}_{50}^\mu/{\rm EC}_{50}^d$ and ${\rm EC}_{50}^\nu/{\rm EC}_{50}^d$ for different values of $\Delta \mu$ and $\Delta \nu$, where $\Delta \mu$ and $\Delta \nu$ are each log-uniformly distributed over a broad range of values (Methods). {\bf (b)} Same as in {\bf (a)}, except now the value of $\Delta \nu$ is constrained so that $\Delta \nu \geq 0.5 \nu_0$.  }
    \label{fig5}
\end{figure}

In the pre-clinical {\em in vitro} or {\em in vivo} setting, the drug effect parameters $\Delta \mu$, $\Delta \nu$, ${\rm EC}_{50}^\mu/{\rm EC}_{50}^d$ and ${\rm EC}_{50}^\nu/{\rm EC}_{50}^d$ that largely determine the optimal drug frequency and dose size can be estimated given sufficient data, and theoretically optimal strategies can be computed (Discussion). In a clinical setting, however, knowledge of these parameters cannot be assumed. In this case, one potential approach is to assign probability distributions to the parameters and find a treatment strategy that performs well on average:
\begin{align} \label{eq:optproblem}
    (c^\ast,\varphi^\ast) = {\rm argmin}_{(c,\varphi)} \E_{\Delta \mu,\Delta \nu,{\rm EC}_{50}^\mu/{\rm EC}_{50}^d,{\rm EC}_{50}^\nu/{\rm EC}_{50}^d}\log\left(\sigma(c,\varphi)/\sigma(c_{\rm max},1)\right).
\end{align}
By taking the logarithm, we are seeking the cyclic strategy that minimizes the geometric mean of the long-run growth rate relative to MTD therapy.
We will refer to this strategy as the BOA (best on average) strategy. Prior knowledge of the parameters can be built into the computation of the expected value when available. For example, if it is known whether ${\rm EC}_{50}^\mu/{\rm EC}_{50}^d$ or ${\rm EC}_{50}^\nu/{\rm EC}_{50}^d$ are more likely to be $>1$ or $<1$, or whether $\Delta \mu$ and $\Delta \nu$ are more likely to be large or small, their distributions can be adjusted to reflect this information. 

We begin by assuming that the parameters $\Delta \mu$, $\Delta \nu$, ${\rm EC}_{50}^\mu/{\rm EC}_{50}^d$ and ${\rm EC}_{50}^\nu/{\rm EC}_{50}^d$  are each log-uniformly distributed over a broad range of values, representing a heterogeneous virtual patient population under minimal prior knowledge of the drug effect parameters (Methods). In this case, the BOA strategy applies a moderate dose ($6.2 \times {\rm EC}_{50}^d$) intermittently (82\% of the time). The effect of this strategy on a representative part of the virtual patient population is visualized in Figure \ref{fig5}a, where blue indicates benefit and red indicates detriment relative to MTD. The BOA strategy benefits approximately 70\% of virtual patients. For these patients, it reduces the long-run growth rate by an average of 6\% and a maximum of 36\% relative to MTD therapy (Table \ref{tab:1}). For the remaining 30\% of patients, the BOA strategy increases the long-run growth rate by an average of 2\% and a maximum of 9\% relative to MTD therapy (Table \ref{tab:1}). Thus, the BOA strategy benefits the majority of patients and the magnitudes of the benefits exceed the magnitudes of the detriments, both in terms of the average effect and the range of observed effects. The prevalence of light colors in Figure \ref{fig5}a indicates that the benefit of the BOA strategy is generally modest, but for a small group of patients (12\%), the growth rate is reduced by over 10\% (Table 1). In addition, for the vast majority of patients, the BOA strategy achieves at least a comparable outcome to MTD therapy using a substantially lower cumulative dose. We stress that the virtual patient population includes both very small and very large drug effects, and still it is possible to identify a single intermittent moderate-dose strategy that improves upon MTD therapy on average across the population, while administering a substantially lower cumulative dose than MTD therapy.

\begin{table}[]
    \hspace*{-60pt}
    \begin{tabular}{|p{4.7cm}|p{1.7cm}|p{1.4cm}|p{1.7cm}|p{1.7cm}|p{1.7cm}|p{2.35cm}|}
        \hline
         Constraints & Patients benefiting & Average benefit & Maximum benefit & Average detriment & Maximum detriment & Patients with $>10\%$ benefit \\
                 \hline
         Baseline & 70.4\%& 5.8\%& 36.0\%& 2.0\%& 8.6\%& 12.1\% \\
                 \hline
          $\Delta \nu \geq 0.5 \nu_0$ & 96.3\%& 16.7\%& 35.8\%& 2.7\%& 10.6\%& 79.1\%\\
                  \hline
        $\qquad \&\; \Delta \mu \leq 10 \mu_0$& 100\%& 18.6\%& 35.3\%&--& --& 97.1\% \\
                \hline
         \raggedright ${\rm EC}_{50}^\mu/{\rm EC}_{50}^d,{\rm EC}_{50}^\nu/{\rm EC}_{50}^d > 1$ & 86.7\% & 9.2\%& 37.5\%& 0.7\%& 8.1\%& 27.9\%\\
                 \hline
       $\qquad \&\; \Delta \nu \geq 0.5 \nu_0$ & 99.9\%& 25.9\%& 37.6\%& 1.3\%& 5.4\%& 98.4\%\\
               \hline
    \end{tabular}
    \caption{
    {\bf Benefit of best-on-average (BOA) strategy relative to MTD therapy in a group of virtual patients.}
    The first column shows the assumptions made on the drug effect parameters $\Delta \mu$, $\Delta \nu$, ${\rm EC}_{50}^\mu/{\rm EC}_{50}^d$ and ${\rm EC}_{50}^\nu/{\rm EC}_{50}^d$ when computing the BOA strategy. The second column shows the proportion of virtual patients whose long-run tumor growth rate is lower under the BOA strategy than under MTD therapy. The third and fourth columns show the average and maximum benefit of the BOA strategy for this group, in terms of percentage decrease in long-run growth rate relative to the MTD therapy. The fifth and sixth columns show the average and maximum detriment for the remaining group of patients, in terms of percentage increase in long-run growth rate relative to the MTD therapy. The final column shows the proportion of virtual patients whose long-run tumor growth rate is more than 10\% lower under the BOA strategy vs.~MTD therapy.}
    \label{tab:1}
\end{table}

Figure \ref{fig5}a indicates that the benefit of the BOA strategy strongly depends on the value of $\Delta \nu$. If we assume that $\Delta \nu$ is known to be large and recompute the BOA strategy given this information, it continues to apply a moderate dose ($6.6 \times {\rm EC}_{50}^d$), but less frequently (64\% of the time). Now, the BOA strategy benefits almost all virtual patients (96.3\%), and for almost 80\% of patients, it reduces the growth rate by over 10\% relative to MTD therapy (Table 1). This is reflected visually by Figure \ref{fig5}b, where the blue color is more prevalent and more intense compared to Figure \ref{fig5}a. If we further assume that $\Delta \mu$ is known to be moderate, the BOA strategy is similar ($7.7 \times {\rm EC}_{50}^d$ applied 59\% of the time), but now it reduces the growth rate by over 10\% in 97\% of cases (Table 1). Thus, even partial knowledge of the maximal drug effect parameters $\Delta \mu$ and $\Delta \nu$ substantially improves BOA performance. 

If it is known that ${\rm EC}_{50}^\mu/{\rm EC}_{50}^d>1$ and ${\rm EC}_{50}^\nu/{\rm EC}_{50}^d>1$, the BOA strategy applies a low dose ($1.9 \times {\rm EC}_{50}^d$) continuously, and it benefits 87\% of patients, with an average detriment of 0.7\% for the remaining patients (Table 1, Figure \ref{fig:s3}). 
If we further assume that $\Delta \nu$ is large, the BOA strategy applies an even lower dose ($1.4 \times {\rm EC}_{50}^d$) continuously, and it now benefits almost all patients ($99.9\%$), with an average growth rate reduction of 26\% (Table 1). 
We know from the previous section that low-dose continuous treatment is optimal when ${\rm EC}_{50}^\mu/{\rm EC}_{50}^d > 1$ and ${\rm EC}_{50}^\nu/{\rm EC}_{50}^d >1$, but we now see that a specific low-dose therapy can be identified that performs well even when the exact parameter values are not known.

The previous paragraphs point to two types of strategies: a fixed moderate dose applied intermittently and a fixed low dose applied continuously. The former strategy is optimal (in the sense of \eqref{eq:optproblem}) if little is known about the underlying parameters, and it becomes even more beneficial if it is known that $\Delta \nu$ is large and $\Delta \mu$ is moderate. The latter strategy is optimal when it is known that ${\rm EC}_{50}^\mu/{\rm EC}_{50}^d >1$ and ${\rm EC}_{50}^\nu/{\rm EC}_{50}^d >1$, and in this case it is  beneficial to a vast majority of patients, especially if it is also known that $\Delta \nu$ is large. 
Overall, these results suggest that the sharp transitions between optimal therapy regimes can be used to reduce the space of optimal strategies to a small number of dose–frequency combinations that perform well even under limited knowledge of the underlying dynamics, while applying a substantially lower cumulative dose than MTD therapy.

\begin{figure}
    \centering
    \hspace*{-48pt} \includegraphics[scale=0.85]{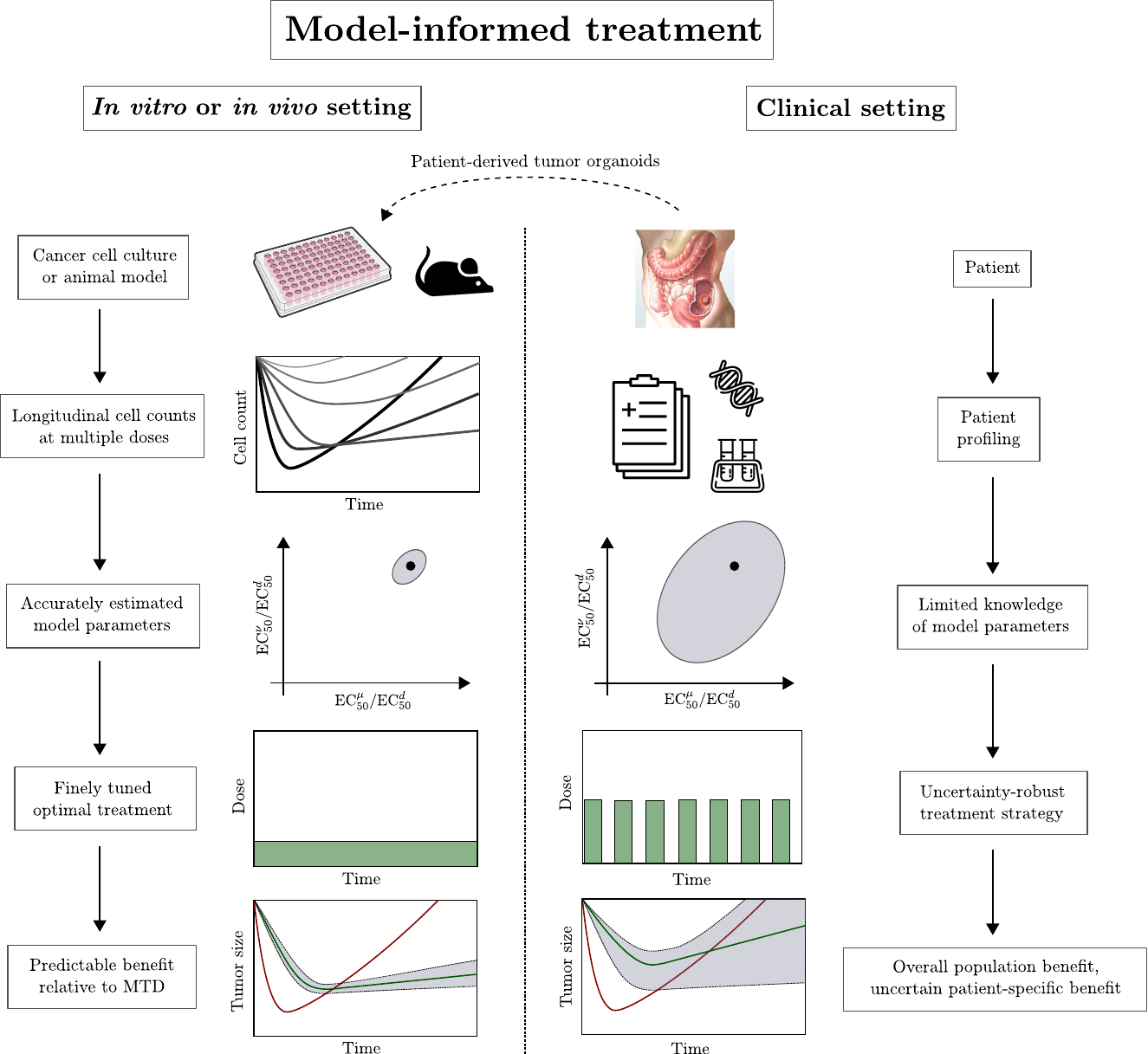}
    \caption{
    {\bf Potential applications of optimal therapy results.}
    Our results support both finely-tuned optimal strategies and 
    robust strategies that are suboptimal yet outperform MTD therapy on average.
    The finely-tuned strategies are applicable to the {\em in vitro} or {\em in vivo} setting where longitudinal data across multiple doses can be collected and model parameters can be estimated.
    Robust strategies may be more suitable when such data are difficult to obtain, as is usually the case clinically.
    The finely-tuned strategies may become more clinically relevant as technologies to derive personalized {\em in vitro} cell models like patient-derived tumor organoids advance.
    }
    \label{fig6}
\end{figure}

\section{Discussion}

Drug-induced persistence is increasingly recognized as an important driver of resistance evolution in cancer.
In this work, we have shown that optimal treatment strategies under drug-induced persistence are characterized by sharp transitions between qualitatively distinct therapy regimes,
governed by the relative dose sensitivities of cell kill, 
tolerance adoption and reversion to sensitivity. 
When the drug inhibits reversion to sensitivity without inducing tolerance adoption, the optimal strategy transitions abruptly from intermittent maximum dosing to continuous low dosing, precisely when the dose sensitivities of cell kill and reversion to sensitivity are equal..
In general, the dose sensitivity of reversion primarily controls the optimal drug frequency while the dose sensitivity of tolerance adoption primarily controls the optimal dose.
In our model, intermittent treatment is optimal only when the drug inhibits reversion at doses that are too low to effectively kill the sensitive cells, 
regardless of whether it also induces tolerance adoption.
We finally showed that the simple regime structure of optimal therapies 
gives rise to
a small number of dose-frequency combinations
that perform well across a broad range of drug effects on phenotypic transitions.

Depending on the setting, our results give rise to finely-tuned optimal strategies or robust strategies that are suboptimal yet outperform MTD therapy on average
(Figure \ref{fig6}). In a pre-clinical {\em in vitro} or {\em in vivo} setting, it is possible to accurately parametrize a mathematical model of drug-induced persistence given sufficient data. For example, Russo et al.~\cite{russo2022modified} and Wu et al.~\cite{wu2025statistical} have shown how the effect of the drug on the transition rate from sensitivity to tolerance can be inferred using longitudinal data on total cell counts across multiple doses. Buder et al.~\cite{buder2017celltrans}, Jagannathan et al.~\cite{jagannathan2020transcompp} and Gunnarsson et al.~\cite{gunnarsson2023statistical} have furthermore shown how the transition rates between states can be determined for individual doses 
when biomarkers can be used to distinguish and separate cells the sensitive and tolerant states. Once the model has been parametrized, the optimal dosing strategy can be computed, which can then guide further pre-clinical or clinical testing for the specific cancer type and anti-cancer drug under investigation. This approach may also eventually support precision medicine as technologies to derive personalized and physiologically relevant
{\em in vitro} cell models like patient-derived tumor organoids advance \cite{wensink2021patient,jiang2020automated, tebon2023drug, walsh2014quantitative,gunnarsson2024understanding}. However, at present, individually tailored optimal strategies are not attainable in the clinical setting, due to the inherent difficulty in collecting longitudinal data on tumor evolution. 
In this case, it may be possible to exploit the simple regime structure of optimal therapies to determine robust dose-frequency combinations that improve upon MTD therapy on average, using experimental and clinical studies involving the same cancer type, as well as treatment histories for similar patients, to inform the model parameter distributions.

Our underlying family of mathematical models is intentionally simple and may require extensions to become more relevant to the clinical setting. 
For example, we have used the long-run tumor growth rate to determine the optimal therapy, which may not reflect optimal therapies under finite treatment windows.
We have also ignored the fact that drug-tolerant cells can develop permanent resistance mechanisms under prolonged drug exposure, through epigenetic reprogramming \cite{shaffer2017rare,francca2024cellular} or genetic mutations \cite{Engelman2016,ramirez2016diverse,russo2022modified}. 
In general, the optimal therapy may change depending on the time scale under consideration.
In addition, the dynamics of tumor evolution may change over time. This suggests the potential need to continuously monitor the tumor and reassess the optimal strategy over time. Model-predictive or receding-horizon control is an example of a framework for continuously updating the model parameters and recomputing the optimal strategy as new data are collected \cite{chen2012optimal,czako2021chemotherapy}. Combining mathematical models with data-driven approaches such as reinforcement learning has also been suggested for this purpose \cite{maier2021reinforcement,gallagher2024mathematical}. The clinical feasibility of model-informed therapies for drug-induced persistence
will likely be closely tied to the identification of DTP biomarkers, which remains an active area of investigation in cancer biology \cite{pu2023drug,russo2024cancer}.
In future work, we plan to incorporate evolution towards more permanent forms of resistance and explore adaptive strategies that are more robust both to model and parameter uncertainty and more responsive to potential changes in the dynamics over time.

\section{Methods}

\subsection{Long-run growth rate under cyclic dosing strategies}

Let $n_0(t)$ be the number of sensitive cells at time $t$ and let $n_1(t)$ be the number of tolerant cells at time $t$.
The time evolution of our model is described by the differential equations
\begin{align*}
\begin{split}
        & \frac{dn_0}{dt} = (\lambda_0(c)-\mu(c))n_0+\nu(c)n_1, \\
    & \frac{dn_1}{dt} = (\lambda_1-\nu(c))n_1+\mu(c)n_0.
\end{split}
\end{align*}
If we let ${\bf n}(t) = [n_0(t), n_1(t)]$ be the (row) vector of subpopulations counts at time $t$, and we define the dose-dependent matrix
\[
{\bf A}(c) = \begin{bmatrix} \lambda_0(c)-\mu(c) & \mu(c) \\ \nu(c) & \lambda_1-\nu(c) \end{bmatrix},
\]
the above system of equations can be written as a matrix differential equation,
\[
\frac{d {\bf n}}{dt} = {\bf n} {\bf A}(c).
\]
If we apply a cyclic strategy where the dose $c$ is applied $\varphi \in [0,1]$ proportion of the time, and we assume that the length of each cycle is arbitrarily short,
we get the differential equation
\[
\frac{d {\bf n}}{dt} = {\bf n} \left(\varphi {\bf A}(c) + (1-\varphi){\bf A}(0)\right),
\]
whose solution can be written using a matrix exponential,
\begin{align*}
    {\bf n}(t) = {\bf n}(0) e^{t \left(\varphi {\bf A}(c) + (1-\varphi){\bf A}(0)\right)}.
\end{align*}
If we define the average rates of $\lambda_0$, $\mu$ and $\nu$ over each treatment cycle,
\begin{align*} 
        & \bar\lambda_0(c,\varphi) := \varphi \lambda_0(c) + (1-\varphi)\lambda_0 = \lambda_0 + \varphi (\lambda_0(c)-\lambda_0), \\
    &  \bar\mu(c,\varphi) := \varphi \mu(c) + (1-\varphi)\mu_0 = \mu_0 + \varphi (\mu(c)-\mu_0), \\
    &  \bar\nu(c,\varphi) := \varphi \nu(c) + (1-\varphi)\nu_0 = \nu_0 + \varphi (\nu(c)-\nu_0),
\end{align*}
then 
\begin{align*}
    \varphi {\bf A}(c) + (1-\varphi){\bf A}(0) = \begin{bmatrix} \bar\lambda_0(c,\varphi)-\bar\mu(c,\varphi) & \bar\mu(c,\varphi) \\ \bar\nu(c,\varphi) & \lambda_1-\bar\nu(c,\varphi) \end{bmatrix} =: {\bf A}(c,\varphi).
\end{align*}
The subpopulation sizes are therefore given by
\begin{align} \label{eq:n(t)expr}
    {\bf n}(t) = {\bf n}(0) e^{t {\bf A}(c,\varphi)}.
\end{align}
Under these dynamics, it is well known that the population eventually grows exponentially, and that the long-run growth rate is given by
\begin{align*}
    \sigma(c,\varphi) :=  \frac12 \Big(&(\bar\lambda_0(c,\varphi)-\bar\mu(c,\varphi)) + (\lambda_1-\bar\nu(c,\varphi)) \nonumber \\
    &+ \sqrt{((\bar\lambda_0(c,\varphi)-\bar\mu(c,\varphi))-(\lambda_1-\bar\nu(c,\varphi)))^2+4\bar\mu(c,\varphi)\bar\nu(c,\varphi)}\Big),
\end{align*}
see Appendix B in \cite{gunnarsson2020understanding}.
In the main text, we also make reference to the recurrence time $\tau$, which is defined as the time it takes the tumor to return to its original size under treatment:
\begin{align*}
    \tau = \inf\{t > 0: n_0(t)+n_1(t) = n_0(0)+n_1(0)\}.
\end{align*}
This quantity is readily computed for a given cyclic strategy using expression \eqref{eq:n(t)expr}.

\subsection{Linear and uniform induction of tolerance}

In our model, the sensitive cell death rate $d_0$ and transition rate $\mu$ from sensitivity to tolerance
are given by the expressions
\begin{align*} 
    & d_0(c) = d_0 + (d_{\rm max}-d_0) \cdot \frac{c}{c+{\rm EC}_{50}^d} = d_0 + \Delta d_0 \cdot \frac{c/{\rm EC}_{50}^d}{c/{\rm EC}_{50}^d+1}, \\
    &\mu(c) = \mu_0 + (\mu_{\rm max}-\mu_0) \cdot \frac{c}{c+{\rm EC}_{50}^\mu} = \mu_0 + \Delta \mu \cdot \frac{c/{\rm EC}_{50}^d}{c/{\rm EC}_{50}^d+{\rm EC}_{50}^\mu/{\rm EC}_{50}^d}.
\end{align*}
If ${\rm EC}_{50}^{\mu} \gg {\rm EC}_{50}^{d}$, then over the dynamic range of the drug, we can assume that $c \ll {\rm EC}_{50}^{\mu}$. In this case, $\mu$ can be approximated as follows using Taylor expansion:
\begin{align*}
    \mu(c) =  \mu_0+\Delta \mu \cdot \frac{{c}/{\rm EC}_{50}^{\mu}}{{c}/{{\rm EC}_{50}^{\mu}}+1} \approx \mu_0 + \frac{\Delta \mu}{{\rm EC}_{50}^\mu} \cdot c,
\end{align*}
meaning that $\mu(c)$ can be approximated as a linear function of the drug dose over the dynamic range of the drug. If on the other hand ${\rm EC}_{50}^{\mu} \ll {\rm EC}_{50}^{d}$, then over the dynamic range of the drug, we can assume that $c \gg {\rm EC}_{50}^{\mu}$. In this case, $\mu$ can be approximated as follows for $c>0$:
\begin{align*}
    \mu(c) =\mu_0+\Delta \mu \cdot \frac{1}{1+{\rm EC}_{50}^{\mu}/c} \approx \mu_0 + \Delta \mu = \mu_{\rm max},
\end{align*}
meaning that $\mu$ is effectively constant in the presence of drug.
This explains why we refer to the case ${\rm EC}_{50}^{\mu} \gg {\rm EC}_{50}^{d}$ as linear induction of tolerance and ${\rm EC}_{50}^{\mu} \ll {\rm EC}_{50}^{d}$ as uniform induction of tolerance.

\subsection{Baseline parameter regime}

Results are shown using a parameter regime inspired by {\em in vitro} experiments using non-small cell lung cancer cells (PC9) treated with EGFR tyrosine kinase inhibitors (erlotinib and gefitinib), specifically Sharma et al.~\cite{sharma2010chromatin} and Hata et al.~\cite{Engelman2016}. Based on these experiments, we take $\lambda_0 = 0.04$, $\lambda_1 = 0.001$, $\Delta d_0 = 0.08$ and $\mu_0 = 0.00004$ (per hour), as further explained in Appendix A of our previous work \cite{gunnarsson2020understanding}. These parameters reflect the fact that the drug-sensitive cells proliferate rapidly in the absence of drug and die rapidly in its presence, and the drug-tolerant cells proliferate slowly both on and off drug. The transition rate from sensitivity to tolerance in the absence of drug is small compared to the net proliferation rate of sensitive cells, which is consistent with epigenetic modifications being retained on the order of $10$-$10^5$ cell divisions \cite{niepel2009non, sigal2006variability, cohen2008dynamic}. We set $\nu_0 = 0.0004$ with the assumption that cells revert from the tolerant state more freely than they adopt it, which is consistent with dynamics between persister cells and normal cells in {\em Escherichia coli} bacterial populations \cite{norman2015stochastic}, and between stem-like and non-stem-like cells in breast cancer \cite{gupta2011stochastic}. 

\begin{table}[h]
\centering
\begin{tabular}{|c|c|c|c|c|c|}
\hline
Parameter&$\lambda_0$ & $\Delta d_0$& $\lambda_1$& $\mu_0$& $\nu_0$ \\
\hline
Value&0.04&0.08&0.001&0.00004&0.0004 \\
\hline
\end{tabular}
\label{table:parameters}
\end{table}

\subsection{Parameter ranges for virtual patients}

In our initial investigation of robust dosing strategies, we assume the parameters ${\rm EC}_{50}^\mu/{\rm EC}_{50}^d$, ${\rm EC}_{50}^\nu/{\rm EC}_{50}^d$, $\Delta \mu$, $\Delta \nu$ are log-uniformly distributed over the following range.
In each case, we discretize the range to decrease computational burden.
\begin{table}[H]
\centering
\begin{tabular}{|c|c|}
\hline
Parameter&Range \\
\hline
${\rm EC}_{50}^\mu/{\rm EC}_{50}^d$&$[10^{-3},10^3]$\\
${\rm EC}_{50}^\nu/{\rm EC}_{50}^d$&$[10^{-3},10^3]$ \\
$\Delta \mu/\mu_0$&$[10^{-3},10^3]$ \\
$\Delta \nu/\nu_0$&$[10^{-3},10^0]$ \\
\hline
\end{tabular}
\label{table:parameters2}
\end{table} 
\noindent These ranges are then modified in further analysis as described in the main text.
To speed up computations, when searching for the optimal dose $c$ and optimal drug frequency $\varphi$, we restrict the search to doses that are $1 \times, 1.1 \times, 1.2 \times, \ldots, 20 \times$ the ${\rm EC}_{50}^d$ dose, and the drug frequencies $50\%,51\%,52\%,...,100\%$. \\

\noindent {\bf Data availability.} 
Data sharing is not applicable to this article as no datasets were generated 
or analyzed during the current study. \\

\noindent {\bf Code availability.} All codes necessary to reproduce the results of this study 
are available in the Github repository \url{https://github.com/egunnars/optimal_therapy_drug_induced_persistence}.  \\

    \noindent {\bf Author contributions.} 
EBG was solely responsible for the conceptualization, methodology, investigation, analysis, software, visualization and manuscript preparation.
\\

\noindent {\bf Competing interests.} The author declares no financial or non-financial competing interests.  \\

\bibliographystyle{unsrt}
\bibliography{epi}

\newpage

\setcounter{page}{1}
\setcounter{section}{0}
\setcounter{equation}{0}

\section*{Supplementary Note 1}

\section{Cytostatic vs.~cytotoxic drug}

In our model, the sensitive cell death rate $d_0$ is assumed to follow a Michaelis-Menten function of the drug dose $c$:
\begin{align*}
    &d_0(c) = d_0+\Delta d_0\cdot\frac{c}{c+{\rm EC}_{50}^d}.
\end{align*}
Then, the net division rate $\lambda_0$ of sensitive cells is given by
\begin{align} \label{eq:lambda0}
    \lambda_0(c) = b_0-d_0(c) = \lambda_0 - \Delta d_0 \cdot\frac{c}{c+{\rm EC}_{50}^d},
\end{align}
where we simply write $\lambda_0$ instead of $\lambda_0(0)$.
If we assume instead that the drug affects the division rate as follows
\begin{align*}
    b_0(c) = b_0 - \Delta b_0 \cdot \frac{c}{c+{\rm EC}_{50}^d},
\end{align*}
without affecting the death rate, 
the net division rate becomes
\begin{align*}
    \lambda_0(c) = b_0(c) - d_0 = \lambda_0 - \Delta b_0 \cdot \frac{c}{c+{\rm EC}_{50}^d},
\end{align*}
which has the same form as expression \eqref{eq:lambda0}.
Thus, our results apply both to cytotoxic and cytostatic drugs, as long as the net division rate of sensitive cells has the form shown in \eqref{eq:lambda0}.

\section{Mathematical analysis}

Here, we provide mathematical analysis in support of the numerical evidence for a phase transition at ${\rm EC}_{50}^\nu/{\rm EC}_{50}^d = 1$ when the drug only inhibits reversion to sensitivity ($\Delta \mu = 0$). In this case, the long-run tumor growth rate under a cyclic strategy $(c,\varphi)$ is given by
\begin{align} \label{eq:sigma(c,phi)suppl}
    \sigma(c,\varphi) =  \frac12 \Big(&(\bar\lambda_0(c,\varphi)-\mu_0) + (\lambda_1-\bar\nu(c,\varphi)) \nonumber \\
    &+ \sqrt{((\bar\lambda_0(c,\varphi)-\mu_0)-(\lambda_1-\bar\nu(c,\varphi)))^2+4\mu_0\bar\nu(c,\varphi)}\Big),
\end{align}
where
\begin{align} \label{eq:d0mu(c,phi)suppl}
\begin{split}
        & \bar\lambda_0(c,\varphi) := \varphi \lambda_0(c) + (1-\varphi)\lambda_0 = \lambda_0 -  (d_{\rm max}-d_0) \cdot \frac{\varphi c}{c+{\rm EC}_{50}^d}, \\
    &  \bar\nu(c,\varphi) := \varphi \nu(c) + (1-\varphi)\nu_0 = \nu_0 - (\nu_0-\nu_{\rm min}) \cdot \frac{\varphi c}{c+{\rm EC}_{50}^\nu}.
\end{split}
\end{align}

\subsection*{Optimal drug frequency $\varphi_\infty$ is independent of ${\rm EC}_{50}^\nu/{\rm EC}_{50}^d$ when $c \to \infty$}
In the limit of a large dose, $c \to \infty$, the expressions in \eqref{eq:d0mu(c,phi)suppl} reduce to
\begin{align} \label{eq:d0mu(phi)suppl}
    \bar\lambda_0(\varphi) = \lambda_0 - \varphi (d_{\rm max}-d_0) \quad\text{and}\quad \bar\nu(\varphi) = \nu_0 - \varphi (\nu_0-\nu_{\rm min}).
\end{align}
Thus, when $c \to \infty$, the long-run growth rate \eqref{eq:sigma(c,phi)suppl} becomes independent of ${\rm EC}_{50}^\nu/{\rm EC}_{50}^d$, meaning that optimal drug frequency $\varphi_\infty$ is independent of ${\rm EC}_{50}^\nu/{\rm EC}_{50}^d$.

\subsection*{A continuum of strategies is optimal when ${\rm EC}_{50}^\nu/{\rm EC}_{50}^d = 1$}

When ${\rm EC}_{50}^\nu/{\rm EC}_{50}^d = 1$, the expressions in \eqref{eq:d0mu(c,phi)suppl} become
\begin{align*}
    \bar\lambda_0(c,\varphi) = \lambda_0 -  (d_{\rm max}-d_0) \cdot \frac{\varphi c}{c+{\rm EC}_{50}^d} \quad\text{and}\quad
    \bar\nu(c,\varphi) =  \nu_0 - (\nu_0-\nu_{\rm min}) \cdot \frac{\varphi c}{c+{\rm EC}_{50}^d}.
\end{align*}
If we define the variable $x = \varphi c/({c+{\rm EC}_{50}^d})$, we can look at the rates in \eqref{eq:d0mu(c,phi)suppl} as one-dimensional functions, 
\begin{align*}
    \bar\lambda_0(x) = \lambda_0 - x (d_{\rm max}-d_0) \quad\text{and}\quad \bar\nu(x) = \nu_0 - x (\nu_0-\nu_{\rm min}),
\end{align*}
which is the same form as in \eqref{eq:d0mu(phi)suppl}. By the definition of $\varphi_\infty$, we know that the long-run growth rate is minimized when $x = \varphi_\infty$, meaning that all strategies $(c,\varphi)$ that lie on the curve
\begin{align} \label{eq:cphicurve}
    \dfrac{\varphi c}{c+{\rm EC}_{50}^d} = \varphi_\infty \quad\text{i.e.}\quad \frac{c}{{\rm EC}_{50}^d} = \frac{\varphi_\infty}{\varphi-\varphi_\infty}.
\end{align}
are optimal. Under continuous drug exposure, $\varphi=1$, the corresponding dose $c_1$ satisfies
\begin{align*}
    c_1/{\rm EC}_{50}^d = \frac{\varphi_\infty}{1-\varphi_\infty}.
\end{align*}

\subsection*{Growth rate on the curve $c/{\rm EC}_{50}^d = {\varphi_\infty}/({\varphi-\varphi_\infty})$ when ${\rm EC}_{50}^\nu/{\rm EC}_{50}^d = 1+\delta$}

On the curve $c/{\rm EC}_{50}^d = {\varphi_\infty}/({\varphi-\varphi_\infty})$ with $\varphi_\infty < \varphi \leq 1$, the rates in \eqref{eq:d0mu(c,phi)suppl} are given by
\begin{align*}
    &\bar\lambda_0 = \lambda_0 -  (d_{\rm max}-d_0) \varphi_\infty, \\
    &\bar\nu(\varphi) =  \nu_0 -(\nu_0-\nu_{\rm min}) \cdot \frac{\varphi_\infty \varphi}{\varphi_\infty+(\varphi-\varphi_\infty)({\rm EC}_{50}^\nu/{\rm EC}_{50}^d)}.
\end{align*}
If ${\rm EC}_{50}^\nu/{\rm EC}_{50}^d = 1+\delta$ and $|\delta|$ is small, we can write
\begin{align*}
   \frac{\varphi_\infty \varphi}{\varphi_\infty+(\varphi-\varphi_\infty)({\rm EC}_{50}^\nu/{\rm EC}_{50}^d)} 
    &= \frac{\varphi_\infty \varphi}{\varphi_\infty+(\varphi-\varphi_\infty)(1+\delta)}  \\
    &= \frac{\varphi_\infty \varphi}{\varphi+(\varphi-\varphi_\infty)\delta} \\
    &= \frac{\varphi_\infty}{1+(1-\varphi_\infty/\varphi) \delta} \\
    &= \varphi_\infty\left(1-(1-\varphi_\infty/\varphi) \delta + O(\delta^2)\right) \\
    &= \varphi_\infty - \varphi_\infty(1-\varphi_\infty/\varphi) \delta + O(\delta^2),
\end{align*}
where we use that $0 \leq 1-\varphi_\infty/\varphi \leq 1$. It follows that
\begin{align*}
    &\bar\nu(\varphi) =  \nu_0 - \varphi_\infty(\nu_0-\nu_{\rm min}) + \varphi_\infty(\nu_0-\nu_{\rm min}) (1-\varphi_\infty/\varphi)\delta + O(\delta^2),
\end{align*}
which by setting $p := \nu_0 - \varphi_\infty(\nu_0-\nu_{\rm min})$ and $q := \varphi_\infty(\nu_0-\nu_{\rm min})$, we can write as
\begin{align*}
    \bar\nu(\varphi) = p + q (1-\varphi_\infty/\varphi) \delta + O(\delta^2).
\end{align*}
We now analyze the long-run growth rate,
\begin{align*} 
    \sigma(c,\varphi) =  \frac12 \Big(&(\bar\lambda_0-\mu_0) + (\lambda_1-\bar\nu(\varphi)) \nonumber \\
    &+ \sqrt{((\bar\lambda_0-\mu_0)-(\lambda_1-\bar\nu(\varphi)))^2+4\mu_0\bar\nu(\varphi)}\Big).
\end{align*}
We first note that
\begin{align*}
    & ((\bar\lambda_0-\mu_0)-(\lambda_1-\bar\nu(\varphi)))^2 \\
    &= ((\bar\lambda_0-\mu_0)-(\lambda_1-p) + q (1-\varphi_\infty/\varphi) \delta + O(\delta^2)))^2 \\
    &= ((\bar\lambda_0-\mu_0)-(\lambda_1-p))^2 + 2q((\bar\lambda_0-\mu_0)-(\lambda_1-p))(1-\varphi_\infty/\varphi) \delta + O(\delta^2)
\end{align*}
and
\begin{align*}
    & 4\mu_0\bar\nu(\varphi) = 4 \mu_0 p + 4 \mu_0 q (1-\varphi_\infty/\varphi) \delta + O(\delta^2)
\end{align*}
which implies
\begin{align*}
    & ((\bar\lambda_0-\mu_0)-(\lambda_1-\bar\nu(\varphi)))^2+4\mu_0\bar\nu(\varphi) \\
    &= \left(((\bar\lambda_0-\mu_0)-(\lambda_1-p))^2 + 4 \mu_0 p\right) + 2q(((\bar\lambda_0-\mu_0)-(\lambda_1-p))+2\mu_0)(1-\varphi_\infty/\varphi) \delta  + O(\delta^2) \\
    &= \left(((\bar\lambda_0-\mu_0)-(\lambda_1-p))^2 + 4 \mu_0 p\right)\left(1 + \frac{2q(((\bar\lambda_0-\mu_0)-(\lambda_1-p))+2\mu_0)}{((\bar\lambda_0-\mu_0)-(\lambda_1-p))^2 + 4 \mu_0 p} (1-\varphi_\infty/\varphi) \delta  + O(\delta^2)\right).
\end{align*}
It follows that
\begin{align*}
    & \sqrt{((\bar\lambda_0-\mu_0)-(\lambda_1-\bar\nu(\varphi)))^2+4\mu_0\bar\nu(\varphi)} \\
    &= \sqrt{((\bar\lambda_0-\mu_0)-(\lambda_1-p))^2 + 4 \mu_0 p} \left( 1 +  \frac{q(((\bar\lambda_0-\mu_0)-(\lambda_1-p))+2\mu_0)}{((\bar\lambda_0-\mu_0)-(\lambda_1-p))^2 + 4 \mu_0 p} (1-\varphi_\infty/\varphi) \delta + O(\delta^2)\right),
\end{align*}
and therefore the long-run growth rate is
\begin{align*} 
    \sigma(c,\varphi) =\; &  \frac12 \Big((\bar\lambda_0-\mu_0) + (\lambda_1-p) + \sqrt{((\bar\lambda_0-\mu_0)-(\lambda_1-p))^2+4\mu_0p}\Big) \\
    &+ \frac{q}2\left(\frac{((\bar\lambda_0-\mu_0)-(\lambda_1-p))+2\mu_0}{\sqrt{((\bar\lambda_0-\mu_0)-(\lambda_1-p))^2 + 4 \mu_0 p}}-1\right) (1-\varphi_\infty/\varphi) \delta + O(\delta^2),
\end{align*}
which is of the form $a+b(1-\varphi_\infty/\varphi)\delta+O(\delta^2)$.
Thus, whenever $b < 0$, the optimal strategy along this curve switches from $\varphi=\varphi_\infty$ to $\varphi=1$ depending on whether $\delta<0$ or $\delta>0$.

We conclude by determining the sign of $b$.
Set $r := (\bar\lambda_0-\mu_0)-(\lambda_1-p)$ and note that
\begin{align*} 
    b &= \frac{q}2\left(\frac{r+2\mu_0}{\sqrt{r^2 + 4 \mu_0 p}}-1\right) 
    = \frac{q}2 \cdot \frac{r+2\mu_0-\sqrt{r^2 + 4 \mu_0 p}}{\sqrt{r^2 + 4 \mu_0 p}} \\
    &= \frac{q}2 \cdot \frac{(r+2\mu_0)^2 - (r^2+4\mu_0p)}{\sqrt{r^2 + 4 \mu_0 p}\left(r+2\mu_0+\sqrt{r^2 + 4 \mu_0 p}\right)} \\
    &= \frac{q}2 \cdot \frac{4\mu_0r+4\mu_0^2-4\mu_0p}{\sqrt{r^2 + 4 \mu_0 p}\left(r+2\mu_0+\sqrt{r^2 + 4 \mu_0 p}\right)} \\
    &= \frac{q}2 \cdot \frac{4\mu_0(r+\mu_0-p)}{\sqrt{r^2 + 4 \mu_0 p}\left(r+2\mu_0+\sqrt{r^2 + 4 \mu_0 p}\right)}.
\end{align*}
Since we assume $\mu_0 > 0$, we know that $r+2\mu_0+\sqrt{r^2 + 4 \mu_0 p} > 0$.
Also, $q = \varphi_\infty(\nu_0-\nu_{\rm min}) > 0$ whenever $\varphi_\infty > 0$.
The sign of $b$ is therefore determined by
\begin{align*}
    r+\mu_0-p &= ((\bar\lambda_0-\mu_0)-(\lambda_1-p)) + \mu_0 - p = \bar{\lambda}_0-\lambda_1.
\end{align*}

\newpage

\setcounter{figure}{0}  

\renewcommand\thefigure{S\arabic{figure}} 

\begin{figure}
    \centering
    \includegraphics[scale=0.55]{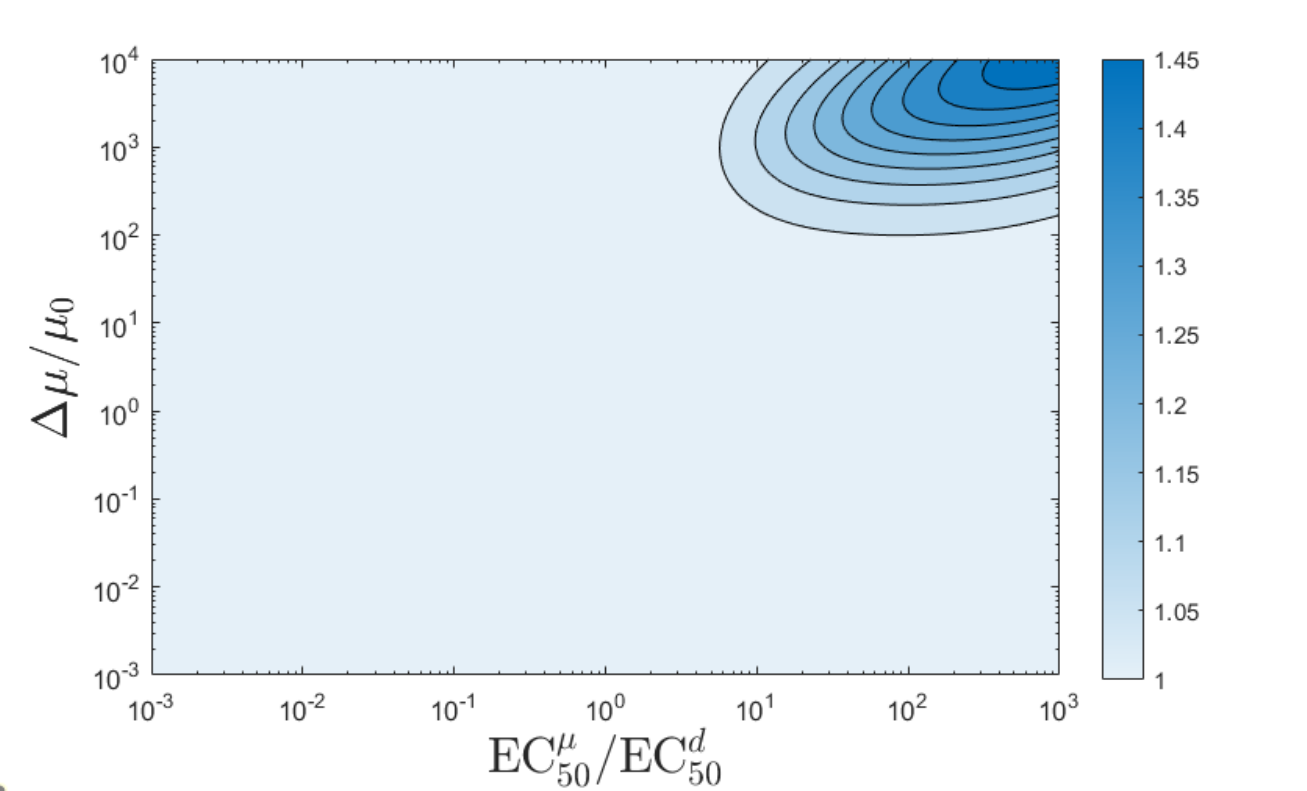}
    \caption{
    {\bf Ratio between long-run growth rate under MTD therapy and the optimal therapy as a function of ${\rm EC}_{50}^\mu/{\rm EC}_{50}^d$ and $\Delta \mu/\mu_0$, when the drug only affects $\mu$.}
    }
    \label{fig:s1}
\end{figure}

\begin{figure}
    \centering
    \includegraphics[scale=0.4]{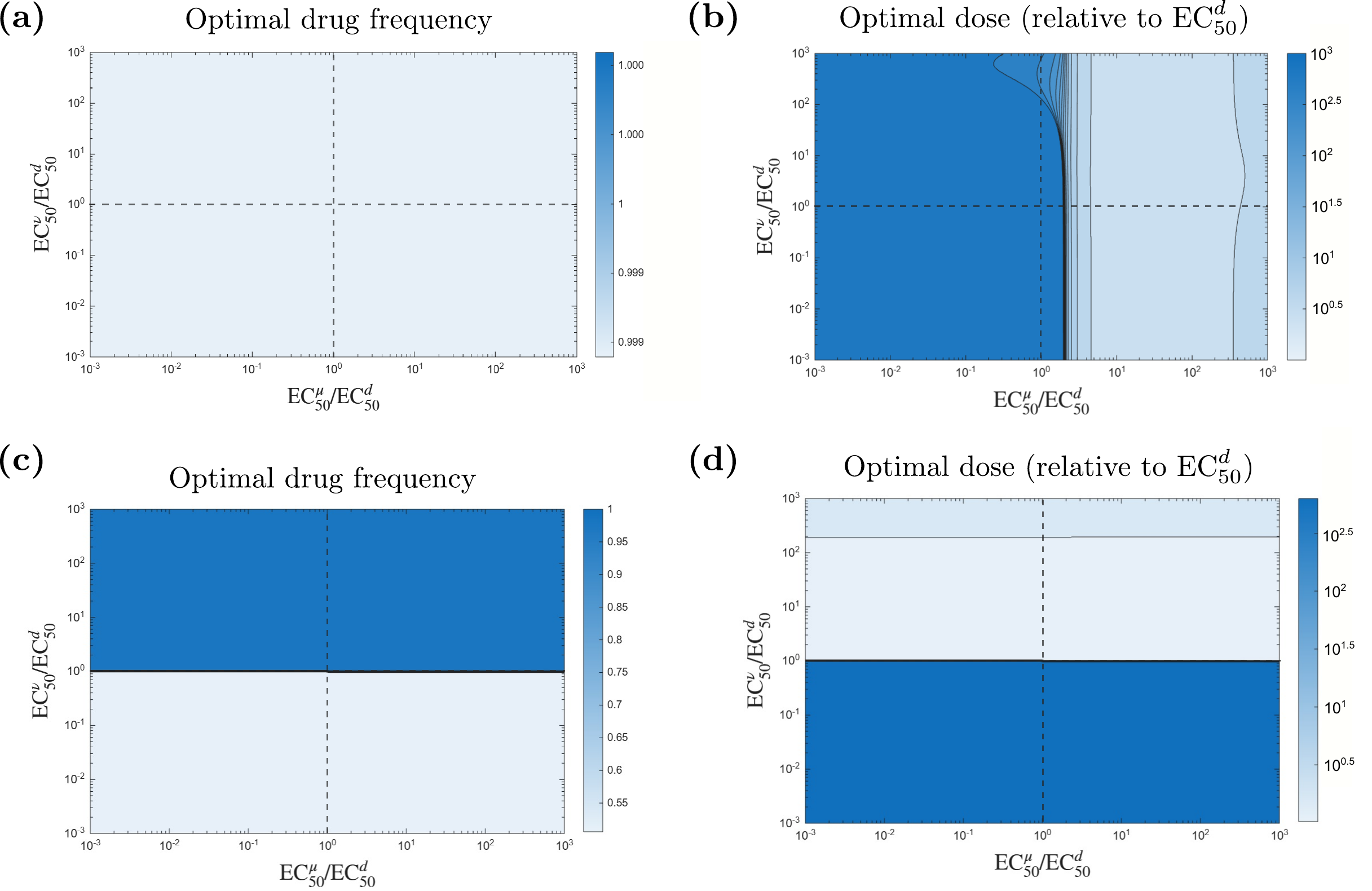}
    \caption{{\bf Optimal therapy when the drug affects both transition rates simultaneously.} {\bf (a)} When $\Delta \nu$ is small ($\Delta\nu/\nu_0 = 10^{-3}$), the optimal therapy is either a continuous low-dose or continuous maximum-dose therapy, as is the case when the drug only affects $\mu$. {\bf (b)} When $\Delta \mu$ is small ($\Delta\mu/\mu_0 = 10^{-2}$), the optimal therapy is either an intermittent maximum-dose or continuous low-dose therapy, as is the case when the drug only affects $\nu$.}
    \label{fig:s2}
\end{figure}

\begin{figure}
\hspace*{-54pt}
    \centering
    \includegraphics[scale=0.43]{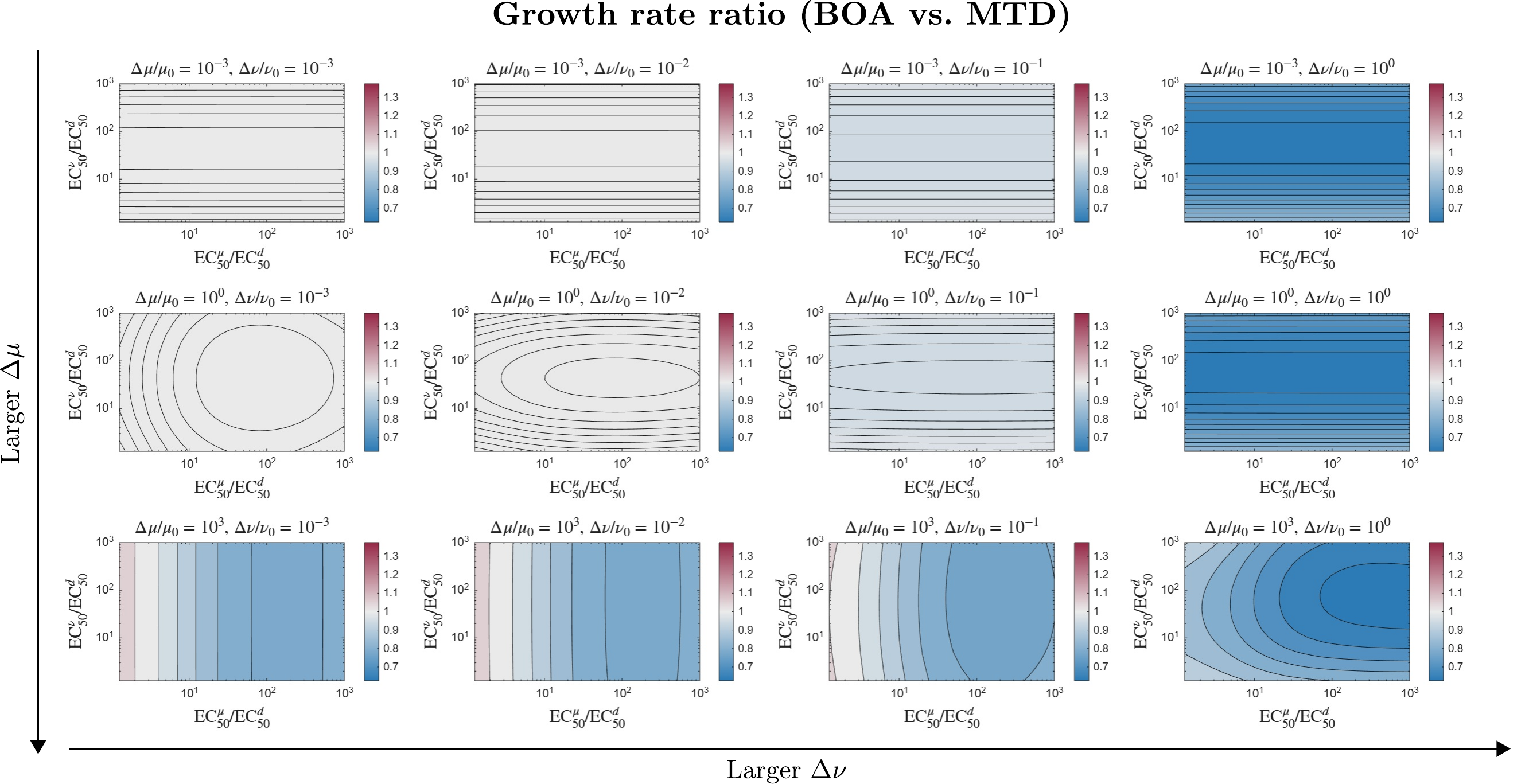}
    \caption{
    {\bf Benefit of best-on-average (BOA) strategy relative to MTD therapy.}
    Each panel shows the ratio between the long-run tumor growth rate under the best-on-average (BOA) strategy vs.~MTD therapy as a function of ${\rm EC}_{50}^\mu/{\rm EC}_{50}^d$ and ${\rm EC}_{50}^\nu/{\rm EC}_{50}^d$ for different values of $\Delta \mu$ and $\Delta \nu$, under the constraints ${\rm EC}_{50}^\mu/{\rm EC}_{50}^d>1$ and ${\rm EC}_{50}^\nu/{\rm EC}_{50}^d>1$.}
    \label{fig:s3}
\end{figure}

\end{document}